\documentclass[10pt,tightenlines,twocolumn]{revtex4}
\usepackage{amsmath,amssymb}
\usepackage[dvips]{graphicx}
\usepackage{psfrag}

\allowdisplaybreaks

\begin{document}

\title{Sub-SQL Sensitivity via Optical Rigidity in Advanced LIGO
Interferometer with Optical Losses }

\author{F. Ya. Khalili, V. I. Lazebny, S. P. Vyatchanin}

\affiliation{Physics Department, Moscow State University,
 Moscow 119992 Russia}

\date{\today}

\begin{abstract}

The ``optical springs'' regime of the signal-recycled configuration of
laser interferometric gravitational-wave detectors is analyzed taking in
account optical losses in the interferometer arm cavities. This regime
allows to obtain sensitivity better than the Standard Quantum Limits both
for a free test mass and for a conventional harmonic oscillator. The optical
losses restrict the gain in sensitivity and achievable signal-to-noise
ratio. Nevertheless, for parameters values planned for the Advanced LIGO
gravitational-wave detector, this restriction is insignificant.

\end{abstract}

\maketitle


\section{Introduction}

Signal-recycled ``optical-springs'' topology of interferometric
gravitational-wave detectors \cite{Buonanno2001, Buonanno2002} is now
considered as a likely candidate to design the second generation of these
detectors, such as Advanced LIGO \cite{WhitePaper1999, Fritschel2002}. This
topology offers an elegant way to overcome the Standard Quantum Limit (SQL)
for a free mass --- a characteristic sensitivity level  when the measurement
noise of the position meter is equal to its back-action noise \cite{67a1eBr,
92BookBrKh, 03a1BrGoKhMaThVy}.

This method is based on the use of optical pondermotive rigidity which
exists in detuned electromagnetic cavities \cite{67a1eBrMa, 70a1eBrMa}. It
turns the test masses in a gravitational-wave detector into harmonic
oscillators, thus providing a resonance gain in test masses displacement
signal \cite{99a1BrKh, 01a1BrKhVo}. In the signal recycling topology, the
pondermotive rigidity can be created relatively easy by adjusting position
of the signal recycling mirror. In this case, only the signal
(anti-symmetric) optical mode eigenfrequency is changed. The arm cavities
and the power (symmetric) mode remain resonance-tuned causing, therefore,
no additional problems with the pumping power.

In large-scale optical systems having bandwidth equal to or  smaller than
the signal frequency $\Omega$, the optical rigidity has  a complicated
frequency dependence \cite{97a1BrGoKh, 01a2Kh, Buonanno2002, 05a1LaVy}.
This  feature allows to  obtain not one but two mechanical resonances and
consequently two minima in  the   noise. Alternatively, these minima can be
placed close  to each other or even superimpose, thus providing a single
wider   ``well'' in the noise spectral density.

The former regime was explored in detail in several articles
\cite{Buonanno2001, Buonanno2002,Buonanno2003,  Harms2003, Buonanno2004,
05a1LaVy}. The latter was examined rather briefly in papers
\cite{01a2Kh, 05a1LaVy}. At the same time, it looks very promising for
detection narrow-band gravitational-wave signals with known frequencies, in
particular, those from the neutron  stars. In this article  we
present an in depth analysis  of this regime.

In Sec.\,\ref{sec:theor_intro} we analyze dynamic properties of the
frequency-dependent pondermotive rigidity in no optical loss case,
including, in particular, an instability inherent to the electromagnetic
rigidity.  This section is based, in part, on the results obtained
in the articles \cite{01a2Kh, Buonanno2003, 05a1LaVy}

In Sec.\,\ref{sec:comp} we compare the optical rigidity-based scheme with
other methods  of circumvent the SQL for a free mass and show that it has a
significant advantage, namely, it is much less vulnerable to optical losses.

In Sec.\,\ref{sec:LIGO} we calculate the sensitivity of the Advanced LIGO
gravitational-wave detectors in the double-resonance optical springs regime.

To clarify this consideration we try to avoid bulky calculations in the main
text. In Appendix \ref{app:analysis} we give detailed  analysis
and calculations of sensitivity and signal-to-noise ratio for Advanced
LIGO interferometer.  In Appendices \ref{app:xi} and \ref{app:snr} we
provide calculations for simplified model without optical losses.

\section{Frequency-dependent optical rigidity.
        No optical losses}\label{sec:theor_intro}

\subsection{``Conventional'' v.s. ``double'' resonances}\label{ord_vs_dbl}

In a single Fabry-Perot cavity, the pondermotive rigidity is equal to
\cite{01a2Kh, 05a1LaVy}:
\begin{equation}\label{K_single}
  K(\Omega) = \frac{2\omega_p{\cal E}}{L^2}\,
    \frac{\delta}{{\cal D}(\Omega)}\,,\quad
    {\cal D}(\Omega) = (-i\Omega+\gamma)^2+\delta^2 \,,
\end{equation}
where  $\Omega$ is the observation (side-band) frequency,
$\omega_p$ is the pumping frequency, $\delta=\omega_p-\omega_o$ is the
detuning, $\omega_o$ is the cavity eigenfrequency, $L$ is its length,
${\cal E}$ is the optical energy stored in the cavity,
and $\gamma$ is the cavity half-bandwidth.

In articles \cite{Buonanno2003, 05a1LaVy} the  signal-recycled
gravitational-wave detectors topology was considered in detail and   its
equivalence to a single cavity was shown. It was shown, in particular, that
Eq.\,(\ref{K_single}) is valid for this  topology too, with obvious
substitution of $\gamma$ and $\delta$ by the anti-symmetric optical mode
half-bandwidth  $\gamma_0$ and detuning $\delta_0$  [see Appendix
\ref{app:analysis} and Eqs.\,(\ref{gamma_0load}),\,(\ref{delta_0})]. Energy
${\cal E}$ in this case is equal to the total optical energy  stored in
both interferometer arms:
\begin{equation}
  {\cal E} = \frac{4I_cL}{c} \,,
\end{equation}
where $I_c$ is the power circulating in each arm.  For the sake
of consistency with other parts of the paper, notations   $\delta_0$
and $\gamma_0$ will be used throughout this paper.

It should be noted that for the narrow-band regimes which
 we dwell on, $\gamma_0$ have to be small:
\begin{equation}\label{nb}
  \gamma_0 \ll \delta_0 \sim \Omega \,.
\end{equation}
Therefore, we neglect for a while term $\gamma_0$, setting
\begin{equation}\label{K_simple}
  K(\Omega) = \frac{K_0\delta_0^2}{\delta_0^2-\Omega^2} \,,
\end{equation}
where $K_0$ is the rigidity value at zero frequency:
\begin{equation}
  K_0 = \frac{2\omega_p{\cal E}}{L^2\delta_0}\,.
\end{equation}
A detailed analysis is provided in \ref{sec:instab} and
\ref{sec:LIGO}.

Consider a harmonic oscillator having mass $m$  and
rigidity (\ref{K_single}). We can write the equation of motion as follows:
\begin{equation}\label{eq_motion}
  m\frac{d^4x(t)}{dt^4} + m\delta_0^2\frac{d^2x(t)}{dt^2}
   + \delta_0^2K_0\,x(t)
   = \frac{d^2F(t)}{dt^2} + \delta_0^2F(t) \,,
\end{equation}
where $x$ is the oscillator coordinate and $F$ is an external force
acting on it.

In general the system has two resonances with frequencies
\begin{equation}\label{Omega_pm}
  \Omega_\pm = \sqrt{\frac{\delta_0^2}{2}
    \pm\sqrt{\frac{\delta_0^4}{4}-\frac{\delta_0^2K_0}{m}}} \,.
\end{equation}
In articles \cite{Buonanno2001, Buonanno2002,
 Buonanno2003} they are
referred to as ``mechanical'' and ``optical'', because in
 asymptotic
case $K_0\to 0$, the frequencies are equal to:
\begin{align}
  \Omega_+ &= \delta_0 \,, & \Omega_- &=
 \sqrt{\frac{K_0}{m}} \,,
\end{align}
Therefore, the high-frequency resonance can be readily
 interpreted as a result
of the optical power sloshing between the optical cavity and
 detuned
pumping field, and the low-frequency one --- as a conventional
resonance of mechanical oscillator with $K_0$ rigidity.

The second-order pole \cite{01a2Kh, 05a1LaVy}, or double
 resonance case takes place
if these two frequencies are equal to each other:
\begin{equation}\label{dbl_res}
  \Omega_+ = \Omega_- = \Omega_0 = \frac{\delta_0}{\sqrt{2}}\,,
\end{equation}
{\it i.e.} if
\begin{equation}\label{K_0}
  K_0 = \frac{m\delta_0^2}{4} \quad\Leftrightarrow\quad
    {\cal E} = {\cal E}_{\rm crit} \,,
\end{equation}
where
\begin{equation}\label{E_crit}
  {\cal E}_{\rm crit} = \frac{mL^2\delta_0^3}{8\omega_0}
\end{equation}
is the critical energy. In this case, the equation of motion
(\ref{eq_motion}) has the following  form:
\begin{equation}\label{eq_motion_double}
  m\left(\frac{d^2}{dt^2} + \Omega_0^2\right)^2x(t)
    = \frac{d^2F(t)}{dt^2} + 2\Omega_0^2F(t) \,.
\end{equation}

It is useful to consider action of resonance force
$F(t)=F_0\cos\Omega_0t$ on this system. Solving
Eq.\,(\ref{eq_motion_double}), we obtain:
\begin{equation}
  x(t) = \frac{F_0}{8m}\left(-t^2\cos\Omega_0t
    + \frac{t\sin\Omega_0t}{\Omega_0}\right) \,.
\end{equation}
The leading term in amplitude of $x(t)$ grows with time as $t^2$. At the
same time, for a conventional oscillator we have
\begin{equation}
  x(t) = \frac{F_0t\sin\Omega_0t}{2m\Omega_0}\,,
\end{equation}
and for a free mass,
\begin{equation}
  x(t) = \frac{F_0\big(1-\cos\Omega_0t\big)}{m\Omega_0^2}\,.
\end{equation}
Therefore, response of the ``double resonance'' oscillator  on the
resonance force is $(\Omega_0t/4)$ times stronger than that of
a conventional harmonic oscillator, and $(\Omega_0t/4)^2$  times stronger
than that of a free mass one.

It was shown in \cite{01a2Kh, 05a1LaVy}, that due to this  feature, the
``double resonance'' oscillator has much smaller value of  the Standard
Quantum Limit for narrow-band signals, than both free  mass and
conventional harmonic oscillators.
It is convenient to express this  gain in
terms of dimensionless parameter
\begin{equation}\label{xi_def}
  \xi^2 = \frac{S_h(\Omega)}{ h_{\rm SQL}^2(\Omega)} \,,
\end{equation}
where $S_h$ is the single-sided spectral density of detector noise
normalized  as the equivalent fluctuational gravitation wave $h$,
\begin{equation}\label{h_SQL}
  h_{\rm SQL}^2(\Omega) = \frac{8\hbar}{mL^2\Omega^2} \end{equation} is
the value of $S_h$ corresponding to SQL. For conventional
SQL-limited gravitation wave detectors (with  free test masses), $\xi\ge
1$.

It is shown in Appendix \ref{app:xi}, that in case of
a conventional first-order resonance, the  equivalent noise curve  has a
``well'' at resonance frequency $\Omega_0$,  which provides the  gain in
sensitivity:
\begin{equation}\label{xi_oscill}
  \xi_{\rm oscill}  = \sqrt{\frac{\Delta\Omega}{\Omega_0}} \,,
\end{equation}
where $\Delta\Omega$ is the bandwidth where this gain is  provided.
In particular, this gain can be obtained by  using either the
``mechanical'' or  ``optical'' resonance of optical rigidity).

In the case of ``double resonance'' oscillator, this gain can  be
substantially more significant:
\begin{equation}\label{xi_dbl}
  \xi_{\rm dbl} = \frac{\Delta\Omega}{\Omega_0}
  = \xi_{\rm oscill}\sqrt{\frac{\Delta\Omega}{\Omega_0}} \,.
\end{equation}
Even better result can be obtained if optical energy is slightly
smaller than the critical value (\ref{E_crit}):
\begin{equation}\label{E_subcrit}
  K_0 = \frac{m\delta_0^2}{4}(1-\eta^2) \quad\Leftrightarrow\quad
    {\cal E} = {\cal E}_{\rm crit}(1-\eta^2)\,, \quad \eta\ll 1\,.
\end{equation}
In this case, function $\xi(\Omega)$ has two minima, which correspond to
two resonance frequencies
\begin{equation}
  \Omega_\pm = \Omega_0\sqrt{1\pm\eta}
  \approx \Omega_0\left(1\pm\frac{\eta}{2}\right) \,,
\end{equation}
and a local maximum at frequency $\Omega_0$. If parameter $\eta$ is  equal
to the optimal value
\begin{equation}
  \eta_c = \xi(\Omega_0) \,,
\end{equation}
 then the bandwidth is $\sqrt{2}$ times wider than in pure double
resonance case for the same value of $\xi$:

\begin{equation}\label{xi_enh_dbl}
  \xi_{\rm enh\,dbl} = \frac{\xi_{\rm dbl}}{\sqrt{2}}
  = \frac{\Delta\Omega}{\sqrt{2}\,\Omega_0}\,.
\end{equation}

Above considerations give an important result
concerning the signal-to-noise ratio values for different regimes \cite{Chen_private}.

The signal-to-noise ratio is equal to:
\begin{equation}\label{snr}
  {\rm SNR} = \frac{2}{\pi}\int_0^{\infty}
    \frac{|h(\Omega)|^2}{S_h(\Omega)}\,d\Omega \,,
\end{equation}
where $h(\Omega)$ is the gravitation wave signal spectrum. It is shown in
Appendix \ref{app:snr}, that for a conventional resonance-tuned
interferometer (without optical springs),
\begin{equation}
  {\rm SNR}_{\rm no\ springs} = {\cal N}\times\frac{|h(\Omega_0)|^2\Omega_0}
    {h^2_{\rm SQL}(\Omega_0)} \,,
\end{equation}
where factor ${\cal N}$ for wide band signal (with bandwidth
$\Delta\Omega_{\rm signal}\ge \Omega_0$) is about unity: for short
pulse ${\cal N}\approx 2.0$ and for step-like signal  ${\cal N}\approx
0.7$. For narrow-band signals with $\Delta\Omega_{\rm
signal}\ll\Omega_0$ we have  ${\cal N} \sim \Delta\Omega_{\rm
signal}/\Omega_0$.

In the narrow-band cases described above the signal-to-noise ratio can be
estimated as follows:
\begin{equation}\label{snr_nb}
  {\rm SNR} \sim \frac{|h(\Omega_0)|^2\Delta\Omega}{S_h(\Omega_0)}
  = \frac{\Delta \Omega}{\xi^2}\,
      \frac{|h(\Omega_0)|^2}{h_{\rm SQL}^2(\Omega_0)} \,.
\end{equation}
(we omit here a numeric factor of the order of unity). In case of a
conventional harmonic oscillator, it can be  shown using
Eqs.\,(\ref{xi_oscill},\ref{snr_nb}), that:
\begin{equation}\label{snr_oscill}
  {\rm SNR}_{\rm oscill}
  \approx \frac{2|h(\Omega_0)|^2\Omega_0}{h_{\rm
 SQL}^2(\Omega_0)}
\end{equation}
(numeric factors in this formula and in Eq.\,(\ref{snr_dbl})  below are
obtained through rigorous integration in Eq.\,(\ref{snr}), see  Appendix
\ref{app:snr}). This value {\em does not depend} on $\Delta\Omega$. Thus
conventional harmonic oscillator provide arbitrary (in the case of zero losses) high
sensitivity at resonance frequency, and also gain in signal-to-noise
ratio equal to $\sim \Omega_0/\Delta\Omega_{\rm signal}$ for narrow-band
signals as compared with conventional interferometer. At the same time, the
value (\ref{snr_oscill}) is fixed and can not be increased by  improving the
meter parameters, and there is no gain in signal-to-noise ratio for
wide-band signals.

On the other hand, in case of double-resonance  oscillator, it
stems form Eqs.\,(\ref{xi_dbl}, \ref{snr_nb}), that:
\begin{equation}\label{snr_dbl}
  {\rm SNR}_{\rm dbl}
  \approx \frac{\sqrt 2 \Omega_0}{\Delta\Omega}\times
  \frac{|h(\Omega_0)|^2\Omega_0}{h_{\rm SQL}^2(\Omega_0)}
  \approx     \frac{ \Omega_0}{\sqrt 2\Delta\Omega} \times
    {\rm SNR}_{\rm oscill}\,.
\end{equation}

Therefore, the double resonance oscillator allows to increase both the resonance frequency
sensitivity and the signal-to-noise ratio by
decreasing the bandwidth $\Delta\Omega$.

The signal-to-noise ratio for Advanced LIGO interferometer with optical
rigidity and influence of optical losses is considered in Sec.\,\ref{sec:snr}.

\subsection{Dynamic instability}\label{sec:instab}

It is well known \cite{67a1eBrMa, 70a1eBrMa} that {\em positive} pondermotive
rigidity is accompanied by {\em negative} dumping, {\it i.e.} a pondermotive
rigidity-based oscillator is always unstable. In case of
frequency-depended rigidity, this instability was calculated in article
\cite{97a1BrGoKh} (see also \cite{Buonanno2001, Buonanno2002}).

If the eigenfrequencies are separated well apart from each other,
$\Omega_+-\Omega_-\gg\gamma_0$, then the characteristic instability time
equals to
\begin{equation}
  \tau_{\rm instab} \approx \left(
    \frac{\gamma_0\Omega_0^2}{\Omega_+^2-\Omega_-^2}
  \right)^{-1}
  \sim \gamma_0^{-1} \sim 0.1\div 1{\rm s}\,.
\end{equation}

The instability becomes more strong, if $\Omega_+-\Omega_-\to
 0$. In
double resonance case ($\Omega_+=\Omega_-$), the instability
 time is equal
to
\begin{equation}
  \tau_{\rm instab}\approx
 \frac{2}{\sqrt{\gamma_0\Omega_0}}\,.
\end{equation}
However, inequality $\Omega_0\tau_{\rm instab}\gg 1$ holds
 in this case too.

It has to be noted that, in principle, any instability can be dumped
without affecting signal-to-noise ratio, if the feedback system sensor
sensitivity is only limited by quantum noises \cite{Buonanno2002} but its
implementation is a separate problem which we do not discuss here..

\section{Comparison with other methods to overcome
SQL and influence of optical losses}\label{sec:comp}

\subsection{``Real'' vs. ``virtual'' rigidities}

From the Quantum Measurements Theory point of view, a laser
 interferometric
gravitational-wave detector can be considered as a meter
 which continuously monitors
position $\hat x(t)$ of a test mass $m$
 \cite{92BookBrKh,
Buonanno2002, 03a1BrGoKhMaThVy}. Output signal of this meter
 is equal to:
\begin{equation}\label{tilde_x}
  \tilde x(t) = \hat x(t) + \hat x_{\rm meter}(t) \,,
\end{equation}
where $\hat x_{\rm meter}(t)$ is the measurement noise and
 $\hat x(t)$ is
the ``real'' position of the test mass. It includes its
 responses on the signal
force
\begin{equation}\label{F_signal}
  F_{\rm signal}(t) = \frac{mL\ddot h(t)}{2}
\end{equation}
and on the meter back-action force $\hat F_{\rm meter}(t)$:
\begin{equation}
  \hat x(t) = {\bf Z}^{-1}[F_{\rm signal}(t) + \hat F_{\rm
 meter}(t)]
\end{equation}
(the terms containing the initial conditions can be omitted, see
 article
\cite{03a1BrGoKhMaThVy}). Here ${\bf Z}$ is a differential
 operator
describing evolution of the test object. For a free test mass
 ({\it i.e.} for
the initial LIGO topology),
\begin{equation}\label{Z_fm}
  {\bf Z} = m\frac{d^2}{dt^2} \,,
\end{equation}
If rigidity $K$ (including the frequency-dependent
 pondermotive one) is
associated with the test mass, then
\begin{equation}\label{Z_oscill}
  {\bf Z} = m\frac{d^2}{dt^2} + K
\end{equation}
Therefore, the signal force estimate is equal to
\begin{equation}\label{tilde_F}
  \tilde F(t) \equiv {\bf Z}\tilde x(t)
  = F_{\rm signal}(t) + \hat F_{\rm sum\,noise}(t)\,,
\end{equation}
where
\begin{equation}\label{F_noise}
  \hat F_{\rm sum\,noise}(t)
  = \hat F_{\rm meter}(t) + {\bf Z}\hat x_{\rm meter}(t)\,.
\end{equation}
is the meter total noise.

The back-action noise $\hat F_{\rm meter}(t)$ is
 proportional to the
amplitude quadrature component of the output light. The
 measurement noise
$\hat x_{\rm meter}(t)$, in the simplest case of SQL-limited
 detector, is
proportional to the phase quadrature component. In this case the spectral
 densities of
these noises satisfy the following uncertainty
 relation (see e.g.
\cite{92BookBrKh}):
\begin{equation}\label{SxSF}
  S_xS_F\ge\hbar^2 \,.
\end{equation}

In more sophisticated schemes which allow to overcome SQL, the
noises $F_{\rm meter}(t)$ and $\hat x_{\rm meter}(t)$ correlate with each
other. In this case, the back-action noise can be presented as follows:
\begin{equation}\label{VM}
  F_{\rm meter}(t) = F_{\rm meter}^{(0)}(t) + {\cal K}\hat x_{\rm meter}(t)
  \,,
\end{equation}
where $F_{\rm meter}^{(0)}(t)$ is the back-action noise component
non-correlated with the measurement noise and coefficient $\cal K$ can be
referred to as ``virtual'' rigidity.

Note that it is precisely the idea of quantum variational measurement
\cite{95a1VyZu, 96a1eVyMa, 96a2eVyMa, 98a1Vy, 02a1KiLeMaThVy}.  In
conventional optical position meters, including the LIGO interferometer, one
measures the phase quadrature component in output wave. This component
contains both the measurement noise ($x_\text{meter}$)  produced by phase
fluctuations in input light wave and the back action noise ($F_{\rm meter}$)
caused by   amplitude fluctuations (optimization of the sum of these two
uncorrelated noises produces SQL). However, using homodyne detector one can
measure  tuned mix of the phase and amplitude quadratures of output waves and
this mix  can be selected in such a way that  the back action noise {\em  can be
compensated} by the noise of the amplitude quadrature. It can be  considered as
introduction of {\em correlation} between the back action and  the measurement noises
as presented in Eq.\,(\ref{VM}). We see that the ``virtual'' rigidity
only relates to the measurement procedure (homodyne angle).

Substituting Eq\, (\ref{VM}) into Eq.\,(\ref{tilde_F}), we obtain, that
\begin{equation}\label{tilde_F_1}
  \hat F_{\rm sum\,noise}(t)
  = \hat F_{\rm meter}^{(0)}(t) + {\bf Z}_{\rm eff}\hat x_{\rm meter}(t)\,,
\end{equation}
where
\begin{equation}\label{two_K}
  {\bf Z}_{\rm eff} = {\bf Z} + {\cal K}
  = m\frac{d^2}{dt^2} + K_{\rm eff} \,,
\end{equation}
and
\begin{equation}
  K_{\rm eff} = K + {\cal K} \,.
\end{equation}
is the {\em effective rigidity}.

Thus, {\em in the lossless case, the total meter noise (\ref{tilde_F_1})
only contains the sum of real rigidity $K$ and virtual one ${\cal K}$, and
 replacement  of any one of them by another one does not change
the total noise spectral density and the signal-to-noise
ratio} \cite{94a1eKhSy}.

Spectral densities of the noises $\hat x_{\rm meter}(t)$ and
 $\hat F_{\rm
fluct}^{(0)}(t)$ also satisfy the uncertainty relation
\begin{equation}\label{SxSF0}
  S_xS_F^{(0)}\ge\hbar^2 \,,
\end{equation}
which does not permit simultaneously making both noise terms in
Eq.\,(\ref{tilde_F_1}) arbitrary small.
 However, factor
${\bf Z}_{\rm eff}$ can be made equal to zero by setting
\begin{equation}\label{K_res}
  K_{\rm eff} = m\Omega^2 \,.
\end{equation}
In this case, only noise $\hat F_{\rm meter}^{(0)}(t)$  remains in
Eq.\,(\ref{tilde_F_1}). In principle this noise can  alone  be made
arbitrary small, thus providing arbitrary high  sensitivity.

Both ``real'' and ``virtual'' (created by the noise  correlation) can be used
for this purposes.  Applying the idea of variational measurement
a simple frequency-independent cross-correlation  (and thus
frequency-independent virtual rigidity ${\cal K}$) can be created relatively
easily by using a homodyne detector. In
this case, Eq.\,(\ref{K_res}) is fulfilled at some given frequency, creating
resonance gain in sensitivity similar to one provided by a conventional
harmonic oscillator.

A frequency-dependent cross-correlation and thus a  frequency-dependent
virtual rigidity ${\cal K}$ can be induced through modification  of the input
and/or output optics of the gravitation-wave detectors using  additional
large-scale filter cavities \cite{02a1KiLeMaThVy}. In this  case,
condition (\ref{K_res}) can be fulfilled and thus the sensitivity better
than SQL is obtained in any given frequency range.

\begin{figure}
  \psfrag{a}{\rm(a)}
  \psfrag{b}{\rm(b)}
  \psfrag{Omega2}[cc][cc]{$(\Omega/\Omega_0)^2$}
  \psfrag{K}[ct][cb]{$K(\Omega)/m\Omega_0^2$}
  \includegraphics[width=0.45\textwidth,height=0.26\textwidth]
    {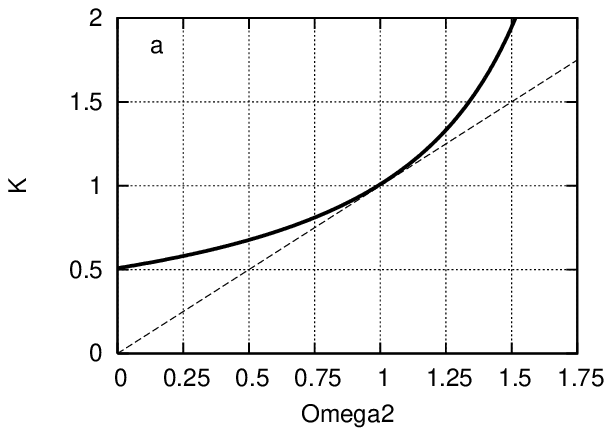}
  \includegraphics[width=0.45\textwidth,height=0.26\textwidth]
    {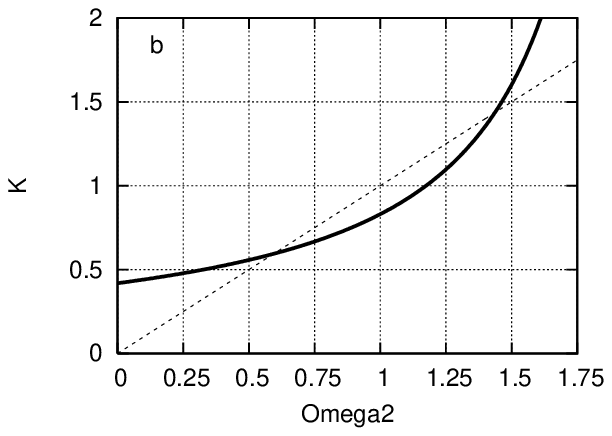}
  \caption{Plots of the frequency-dependent rigidity $K(\Omega)$ as a
  function of $\Omega^2$ (solid line); dashed line corresponds to
  $m\Omega^2$; (a) --- critical pumping (the double resonance  regime); (b)
  --- sub critical pumping.}\label{fig:regimes}
\end{figure}

Consider now the real pondermotive rigidity. It is evident that it can not
be tuned in  such a flexible way  as the virtual one. In the
double-resonance case, graphics of $K(\Omega)$ and $m\Omega^2$
touch each other only at frequency $\Omega_0$.  However, around this
point,
\begin{equation}
\label{Kcross}
  \big(K_{\rm eff}(\Omega)\big) - m\Omega^2 \sim
 (\Omega-\Omega_0)^2 \,,
\end{equation}
instead of $\Omega-\Omega_0$ for the case of a conventional
frequency-independent rigidity. As it was shown in
Sec.\,\ref{sec:theor_intro}, slightly better results can be obtained by
using sub-critical pumping ${\cal E}<{\cal  E}_{\rm crit}$, which provide
two closely placed first-order resonances, see Fig.\,\ref{fig:regimes}(b).

\subsection{Influence of optical losses}\label{sec:losses}

It is possible to conclude from above considerations that the virtual
rigidity provides a more promising solution than the real pondermotive one. However,
as we show in this subsection, the pondermotive rigidity has one
important advantage: it is much less vulnerable to losses in optical
elements. Our consideration will be based on the following statement, which
follows from Eqs.\,(\ref{ae_to_pq}): {\em a lossy optical position meter is
equivalent to the similar lossless one with gray filter attached to its signal
port.} This filter transmittance has  to be equal to:
\begin{equation}
  T_{\rm equiv}^2 = \frac{\gamma_0^{\rm load}}{\gamma_0}\,,
\end{equation}
where
\begin{equation}
\label{gamma0}
  \gamma_0 = \gamma_0^{\rm load} + \gamma_0^{\rm loss}\,,
\end{equation}
$\gamma_0^{\rm load}$ is the term describing signal mode half-bandwidth $\gamma_0$
describing coupling with photodetector and $\gamma_0^{\rm loss}$ is
the term describing optical losses.

Term ``similar'' means, that (i) both meters have same dynamic parameters,
{\it i.e.} bandwidths and eigenfrequencies of the cavities, etc. In
particular, half-bandwidth of the lossless meter has to be equal to
$\gamma_0$ and (ii) optical energies stored in the lossless meter and the
lossy one have to be equal to each other.

Therefore, the lossy meter output signal can be presented in
 the following form
(compare with Eq.\,(\ref{tilde_x})):
\begin{equation}
  \tilde x_{\rm lossy}(t) = \hat x(t) + \hat x_{\rm
 meter}(t)
    + |{\cal A}_0|\hat x_{\rm loss}(t) \,,
\end{equation}
where
\begin{equation}
\label{A0}
  |{\cal A}_0| = \frac{\sqrt{1-T_{\rm equiv}^2}}{T_{\rm
 equiv}}
  = \sqrt{\frac{\gamma_0^{\rm loss}}{\gamma_0^{\rm load}}}
\end{equation}
is the effective loss factor and $\hat x_{\rm loss}(t)$ is  the
additional noise produced by the optical losses uncorrelated with
$x_\text{meter}$. Spectral density of this noise is also equal to $S_x$,
because both $\hat x_{\rm loss}(t)$ and  $\hat x_{\rm loss}(t)$ are in
essence zero-point fluctuations normalized  by the transfer function of
measurement system. The key point is that  while the back-action
noise $\hat F_{\rm meter}(t)$ can correlate with  the measurement noise
$\hat x_{\rm meter}(t)$, it {\em can not  correlate} with additional
noise $\hat x_{\rm loss}(t)$:
\begin{equation}\label{tilde_F_loss}
  \hat F_{\rm sum\,noise}(t) = \hat F_{\rm meter}^{(0)}(t)
    + {\bf Z}_{\rm eff}\hat x_{\rm meter}(t)
    + |{\cal A}_0|{\bf Z}\hat x_{\rm loss}(t)
\end{equation}
[compare with Eq.\,(\ref{tilde_F_1})].

It follows from this equation that in the lossy meter case  there is no
symmetry of the real and virtual rigidities: the term  proportional to
the meter noise $\hat x_{\rm meter}(t)$ still depends on the  effective
rigidity, but the new losses term only depends on the real  rigidity.

Compare now too particular cases of ``pure real'' and ``pure  virtual''
rigidities.

In the first case, ${\bf Z}_{\rm eff}={\bf Z}$, and
\begin{equation}
  \hat F_{\rm sum\,noise}(t) = \hat F_{\rm meter}^{(0)}(t)
    + {\bf Z}\big[\hat x_{\rm meter}(t)
    + |{\cal A}_0|\hat x_{\rm loss}(t)\big]\,.
\end{equation}
Therefore, both terms proportional to $\hat x_{\rm meter}(t)$ and  $\hat
x_{\rm loss}(t)$ can be canceled by setting ${\bf Z}=0$, thus
providing arbitrary high sensitivity at least for one given frequency.

In the second case,
\begin{equation}
  \hat F_{\rm sum\,noise}(t) = \hat F_{\rm meter}^{(0)}(t)
    + {\bf Z}_{\rm eff}\hat x_{\rm meter}(t)
    - m\Omega^2|{\cal A}_0|\hat x_{\rm loss}(t)\,.
\end{equation}
Suppose that ${\bf Z}_{\rm eff}$ is canceled using some QND procedure.
In this case, the sum noise will still consist of two  non-correlated
parts proportional to $\hat F_{\rm meter}^{(0)}(t)$ and $\hat  x_{\rm
loss}(t)$, and its spectral density will be equal to:
\begin{equation}
  S_{\rm sum}(\Omega) = S_F^{(0)} + m^2\Omega^4|{\cal A}_0|^2S_x \,.
\end{equation}
Taking into account uncertainty relation (\ref{SxSF0}), it is easy
to see that
\begin{equation}
  S_{\rm sum}(\Omega) \ge 2|{\cal A}_0|\hbar m\Omega^2 \,,
\end{equation}
or
\begin{equation}\label{xi_min_loss}
  \xi \equiv \sqrt{\frac{S_{\rm sum}(\Omega)}{S_{\rm SQL}(\Omega)}}
  \ge \sqrt{|{\cal A}_0|}
  = \left(\frac{\gamma_0^{\rm loss}}{\gamma_0^{\rm
    load}}\right)^{1/4}\,,
\end{equation}
where
\begin{equation}\label{S_SQL}
  S_{\rm SQL}(\Omega) = \frac{m^2L^2\Omega^4}{4}\,h_{\rm   SQL}^2(\Omega)
  = 2\hbar m\Omega^2
\end{equation}
[see Eqs.\,(\ref{h_SQL},\ref{F_signal})].

The restriction (\ref{xi_min_loss}) shows that the use of ``pure virtual''
rigidity (variational measurement) is very sensitive to losses --- for parameters
from Table \ref{tab1} we have
$\big(\gamma_0^\text{loss}/\gamma_0^\text{load}\big)^{1/4}\simeq 0.7$ (!).

For conventional (resonance-tuned) Advanced LIGO topology,
$\gamma_0\simeq\Omega\simeq 2\pi\times 100{\rm s}^{-1}\gg \gamma_0^{\rm
loss}$ and ``virtual'' rigidity can be introduced by means of the
variational measurement. In this case the optical losses restrict the
sensitivity by the following value:
\begin{equation}\label{xi_min_std}
  \xi\simeq \left(\frac{\gamma_0^{\rm loss}}{\Omega}\right)^{1/4}
  \simeq 0.2
\end{equation}
(for the value of $\gamma_0^{\rm loss}$, see Table\,\ref{tab1}).

In  general both real and virtual rigidities exist  in the
signal-recycled configuration of the interferometric gravitation-wave
detectors as well as for that of a single detuned cavity. However, in the
narrow-band regimes (\ref{nb}), the real rigidity dominates: it is
approximately $\delta_0/\gamma_0$ times stronger than the virtual one
(compare Eqs.\,(\ref{K})) and (\ref{K_eff}), or (\ref{Z_approx}) and
(\ref{Z_approx_eff}), which differ by terms of the order of magnitude
$\lesssim\gamma_0/\delta_0$ only). Due to this reason, this  regime is free
from limitation (\ref{xi_min_loss}).

It is also interesting to note that while both $K$ and  ${\cal K}$
contain, in general, imaginary parts, in effective rigidity  $K_{\rm
eff}$ these imaginary parts exactly compensate each other.

\begin{table}
\caption{Parameters planned to use in Advanced LIGO
 \cite{WhitePaper1999}.
In estimates of $|{\cal A}_0|^2$ and $\gamma_0$
using formulas (\ref{gamma0},\ref{A0},\ref{gamma_0load}) we
assume that $|1+R_se^{2i\phi}|^2\simeq 2$ }\label{tab1}
\begin{tabular}{|p{0.3\textwidth}  p{0.18\textwidth}|}
\hline
Transmissivity of SR mirror&\ $T_s^2=0.05$\\
Transmissivity of input mirrors in arms &\ $T^2=0.005$\\
Loss coefficient of each mirror in arms&\ $ {\cal
 A}_1^2={\cal A}_2^2=
        1.5\times 10^{-5}$\\
Length of interferometer arm& \ $L=4$~km\\
Effective loss factor &\   $|{\cal A}_0|^2= 0.24$\\
Relaxation rate of difference mode
        &\ $\gamma_{0}=2.9\ \text{s}^{-1}$ \\
``Intrinsic'' relaxation rate &\ $\gamma_0^\text{loss} =
 0.56\ \text{s}^{-1}$\\
Mean frequency of gravitational wave range &
       \ $\Omega_{0}=2\pi\times 100\ \text{s}^{-1}$\\
\hline
\end{tabular}
\end{table}

\section{Advanced LIGO sensitivity in double-resonance
regime}\label{sec:LIGO}

\subsection{ The sum noise spectral density}

Sensitivity of the signal-recycled Advanced LIGO topology in the
narrow-band approximation (close to the double resonance) is calculated in
Appendix \ref{app:analysis}. It is shown, that in this case [see
Eq.~(\ref{xi(nu)})]
\begin{equation}
\label{xi_real}
  \xi^2(\Omega) \approx  \xi_0^2+
    \frac{|{\cal A}_\alpha|^2}{4\gamma_0^{\rm loss}\Omega_0^3}
      \left(4\nu^2-\eta_\alpha^2\Omega_0^2\right)^2
\end{equation}
where $\nu=\Omega-\Omega_0$,
\begin{equation}
\label{C}
\xi_0^2=\frac{\gamma_0^{\rm loss}}{\Omega_0|{\cal A}_\alpha|^2}\,C\,,\quad
  C = 1 + \frac{3}{\sqrt{2}}\,|{\cal A}_\alpha|^2 + |{\cal A}_\alpha|^4\,,
\end{equation}
and $\eta_\alpha,\ |{\cal A}_\alpha|$ are  basically parameters $\eta,\
|{\cal A}_0|$ corrected to take into  account virtual
rigidity ({\it i.e.} homodyne angle  $\alpha$), see
Eqs.\,(\ref{eta_A_alpha}).

\begin{figure}
  \psfrag{.1}[rc][cl]{$0.1$}
  \psfrag{1.}[ct][cb]{$1$}
  \psfrag{.1e2}{$10$}
  \psfrag{.1e\2611}{$0.01$}
  \psfrag{.4}[ct][cb]{$0.4$}
  \psfrag{.6}[ct][cb]{$0.6$}
  \psfrag{.8}[ct][cb]{$0.8$}
  \psfrag{2.}[ct][cb]{$2$}
  \psfrag{x}[lt][rb]{$\Omega/\Omega_{00}$}
  \psfrag{y}[cc][lc]{$\frac{\sqrt S_h}{h_{SQL}}$}
  \psfrag{z}[cc][lc]{$\frac{Z(\Omega)}{m\Omega_{00}^2}$}
  \psfrag{alpha=0}[cb][ct]{$\alpha=0$,\quad $\eta=0$,\quad
 $\gamma_{0}/\Omega_0=0.01$}
  \psfrag{alpha=Pi/2}[cb][ct]{$\alpha=\pi/2$,\quad
 $\eta=0$,\quad
    $\gamma_{0}/\Omega_0=0.01$}
  \psfrag{Imp}[cb][ct]{$\eta=0$,\quad
 $\gamma_{0}/\Omega_0=0.01$}
  \vspace{5mm}
  \includegraphics[width=0.45\textwidth,
      height=0.25\textwidth]{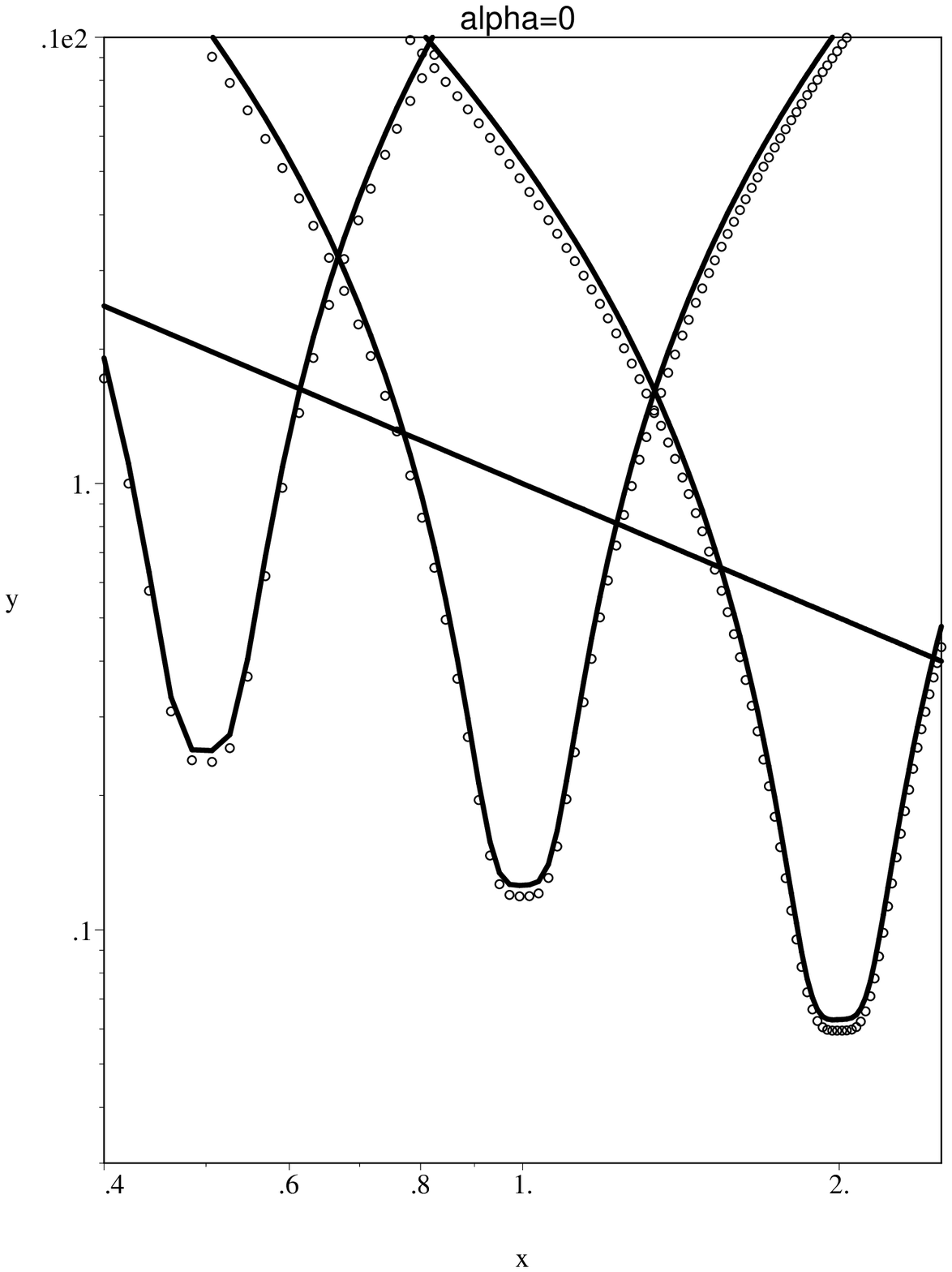}\\[8mm]
  \includegraphics[width=0.45\textwidth,
      height=0.25\textwidth]{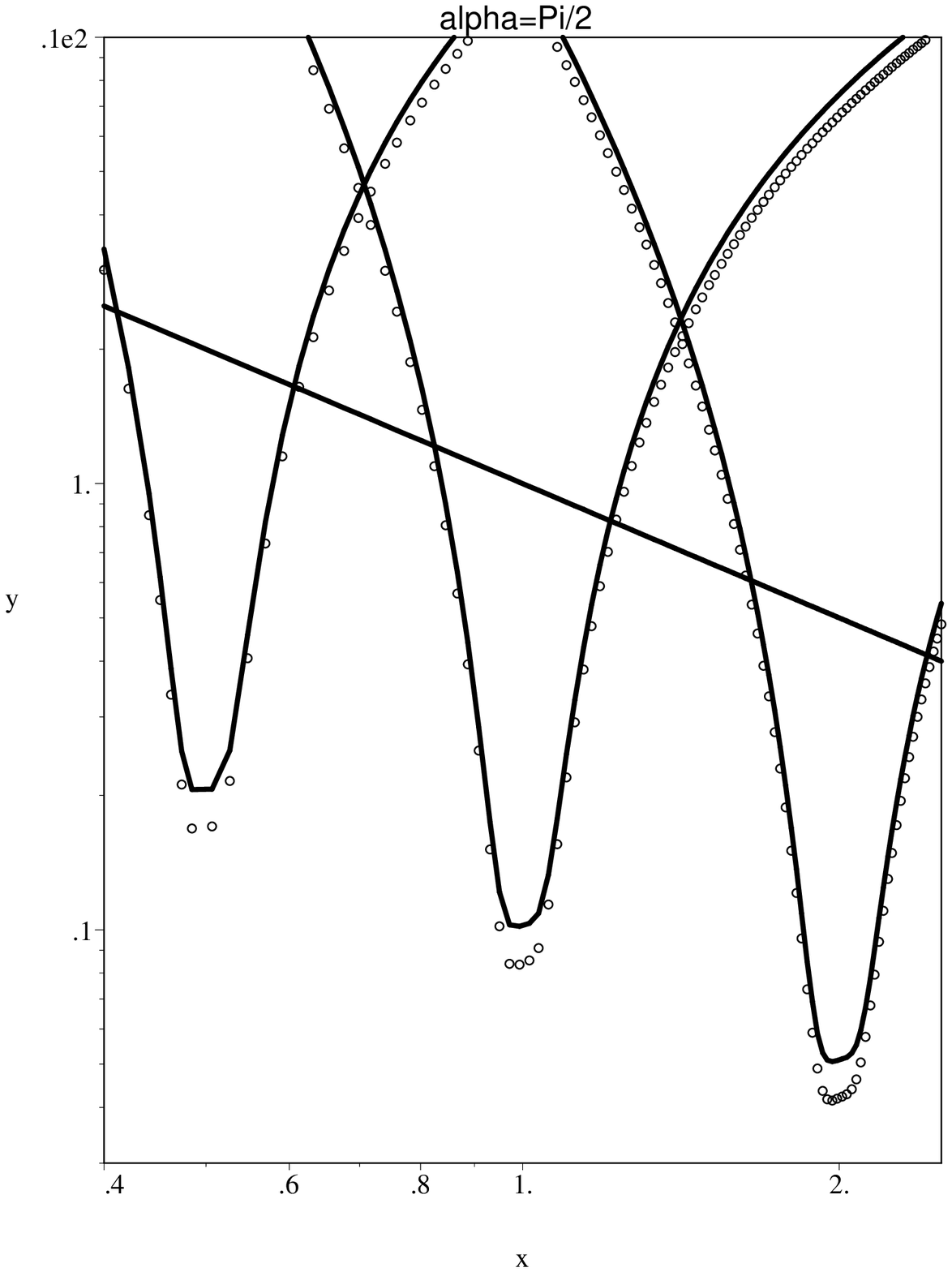}\\[8mm]
  \caption{Top plot: function $\sqrt{S_h(\Omega)}/h_{SQL}(\Omega_{00})$
    plotted for parameters $\alpha=0$, $\gamma_{0}/\Omega_0=.01$,
    $|{\cal A}_0|^2=0.24$ and $\eta_\alpha=0$. The left solid curve ---
    $\Omega_0=0.5\times \Omega_{00}$, the middle solid curve ---
    $\Omega_0=\Omega_{00}$ and the right solid  curve ---
    $\Omega_0=2\times\Omega_{00}$. The dotted curves --- sensitivity
    for no loss case with $\gamma_0^\text{no losses}=\gamma_{0}$.
    Straight line depicts SQL sensitivity. Bottom plot: the
    same as on top plot with homodyne angle $\alpha=\pi/2$.
    $\Omega_{00}$ is some arbitrary frequency, for example,
    $\Omega_{00}=2\pi\times100{\rm s}^{-1}$.
  }\label{oneA}
\end{figure}

\begin{figure}
  \psfrag{.1}[rc][cl]{$0.1$}
  \psfrag{1.}[ct][cb]{$1$}
  \psfrag{.1e2}[cc][lc]{$10$}
  \psfrag{.1e\2611}{$0.01$}
  \psfrag{.7}[ct][cb]{$0.7$}
  \psfrag{.9}[ct][cb]{$0.9$}
  \psfrag{.8}[ct][cb]{$0.8$}
  \psfrag{2.}[ct][cb]{$2$}
  \psfrag{x}[lt][rb]{$\Omega/\Omega_{0}$}
  \psfrag{y}[cc][lc]{$\xi$}
  \psfrag{z}[cc][lc]{$\frac{Z(\Omega)}{m\Omega_{oo}^2}$}
  \psfrag{alpha=0}[cb][ct]{$\alpha=0$,\quad $\eta=0$}
  \psfrag{alpha=Pi/2}[cb][ct]{$\alpha=\pi/2$,\quad $\eta=0$}
  \psfrag{Imp}{}
  \vspace{5mm}
  \includegraphics[width=0.22\textwidth,
    height=0.2\textwidth]{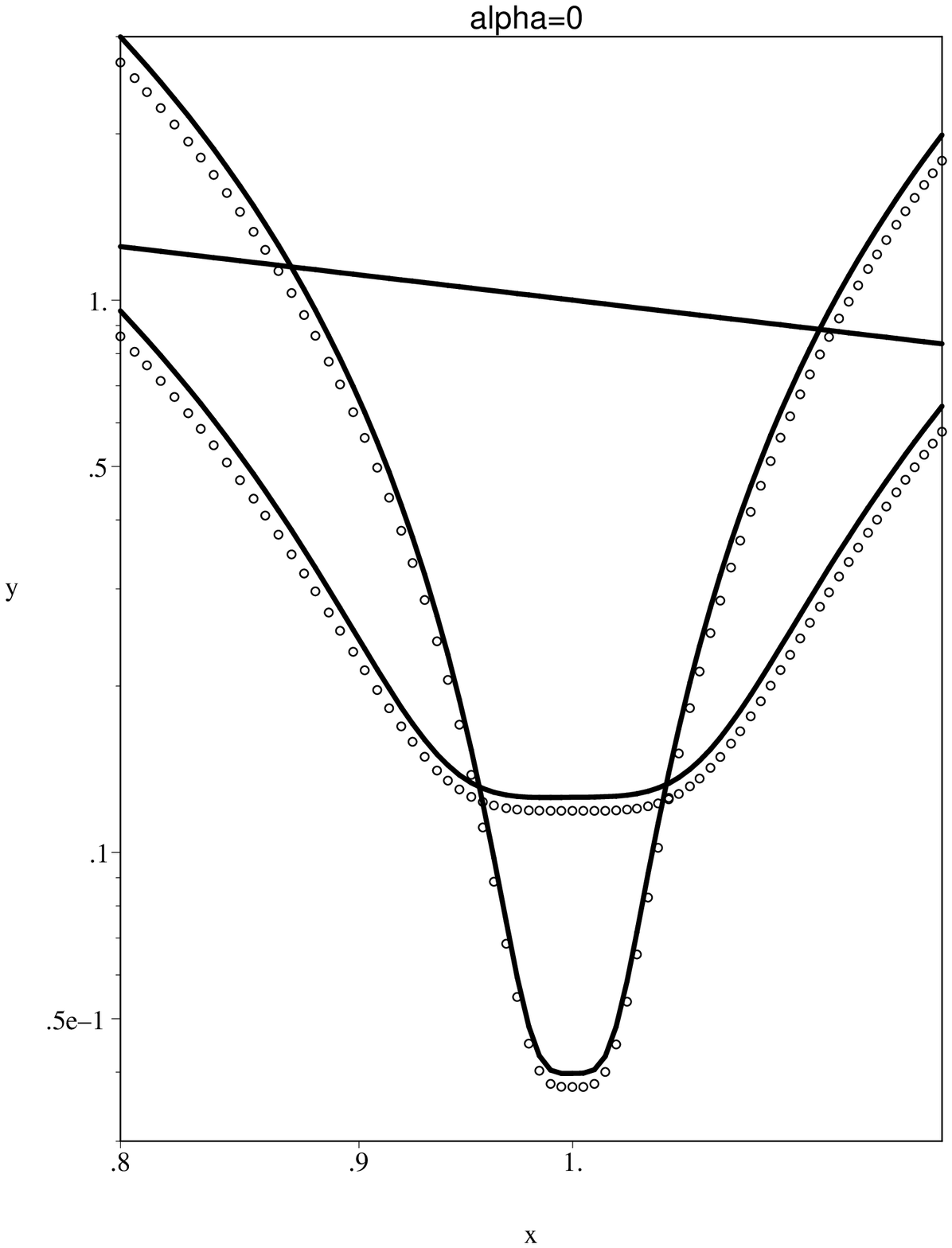} \quad
  \includegraphics[width=0.22\textwidth,
      height=0.2\textwidth]{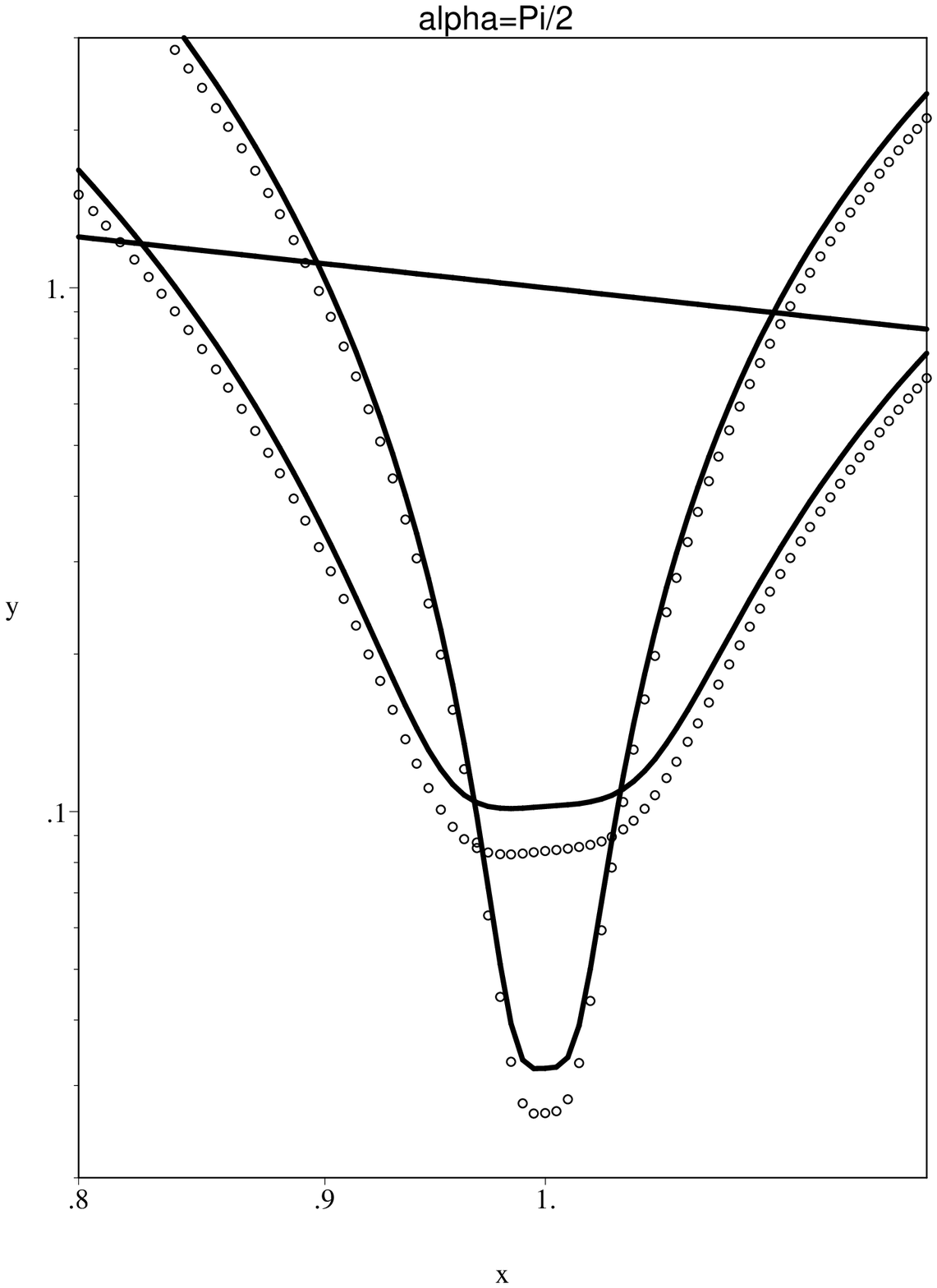}
  \caption{Left plot: sensitivity curves
    $\xi(\Omega)$ plotted for parameters
    $\alpha=0$, $\Omega_0=\Omega_0$, $|{\cal A}_o|^2=0.24$ and
 $\eta=0$. The wider
   solid curve corresponds to $\gamma_{0}/\Omega_0=.01$,
 the narrower
   solid curve --- $\gamma_{0}/\Omega_0=.001$. The dotted
 curves ---
   sensitivity for no loss case with
   $\gamma_0^\text{no losses}=\gamma_{0}$. Straight line
 depicts SQL
   sensitivity. Right plot: the same as on the left plot for
 homodyne angle
   $\alpha=\pi/2$.
 }\label{oneB}

\end{figure}

It follows from Eq.\,(\ref{xi_real}) that the single minimum dependence of
sensitivity $\xi(\Omega)$ takes place if $\eta_\alpha=0$ with bandwidth
$\Delta \Omega$ equal to
\begin{equation}
\Delta \Omega=\sqrt{2\gamma_0^\text{loss}\Omega_0
        \frac{\sqrt C}{|{\cal A}_\alpha|^2}\, }.
\end{equation}
On Figs.\,\ref{oneA}, \ref{oneB} we present the sensitivity plots for various
sets of parameters which allows to realize single minimum dependence of $\xi$
on different frequencies $\Omega_0$ and with different depths for the case
when  effective loss factor $|{\cal A}_0|^2=0.24$ (as planned in Advanced
LIGO, solid curves) and for no losses case (dotted curves). We see that the
sensitivity degradation due to the optical losses is negligibly small.
Possibility of scaling the frequency $\Omega_0$ is  also demonstrated in this
plots.

\begin{figure}
\psfrag{.1}[rc][cl]{$0.1$}
\psfrag{.5}[rc][cl]{}
\psfrag{1.}[ct][cb]{$1$}
\psfrag{5.}[ct][lb]{$5$}
\psfrag{.1e2}{$10$}
\psfrag{.1e\2611}{$0.01$}
\psfrag{.4}[ct][cb]{$0.4$}
\psfrag{.6}[ct][cb]{$0.6$}
\psfrag{.7}[ct][cb]{$0.7$}
\psfrag{.8}[ct][cb]{$0.8$}
\psfrag{2.}[ct][cb]{$2$}
\psfrag{x}[lt][rb]{$\Omega/\Omega_{0}$}
\psfrag{y}[cc][lc]{$\xi$}
\psfrag{z}[cc][lc]{$\frac{Z(\Omega)}{m\Omega_{oo}^2}$}
\psfrag{alpha=0}[cb][ct]{$\alpha=0,\quad \eta=\eta_c$}
\psfrag{alpha=Pi/2}[cb][ct]{$\alpha=\pi/2,\quad
 \eta=\eta_c,$}
\vspace{5mm}
\includegraphics[width=0.23\textwidth,
    height=0.2\textwidth]{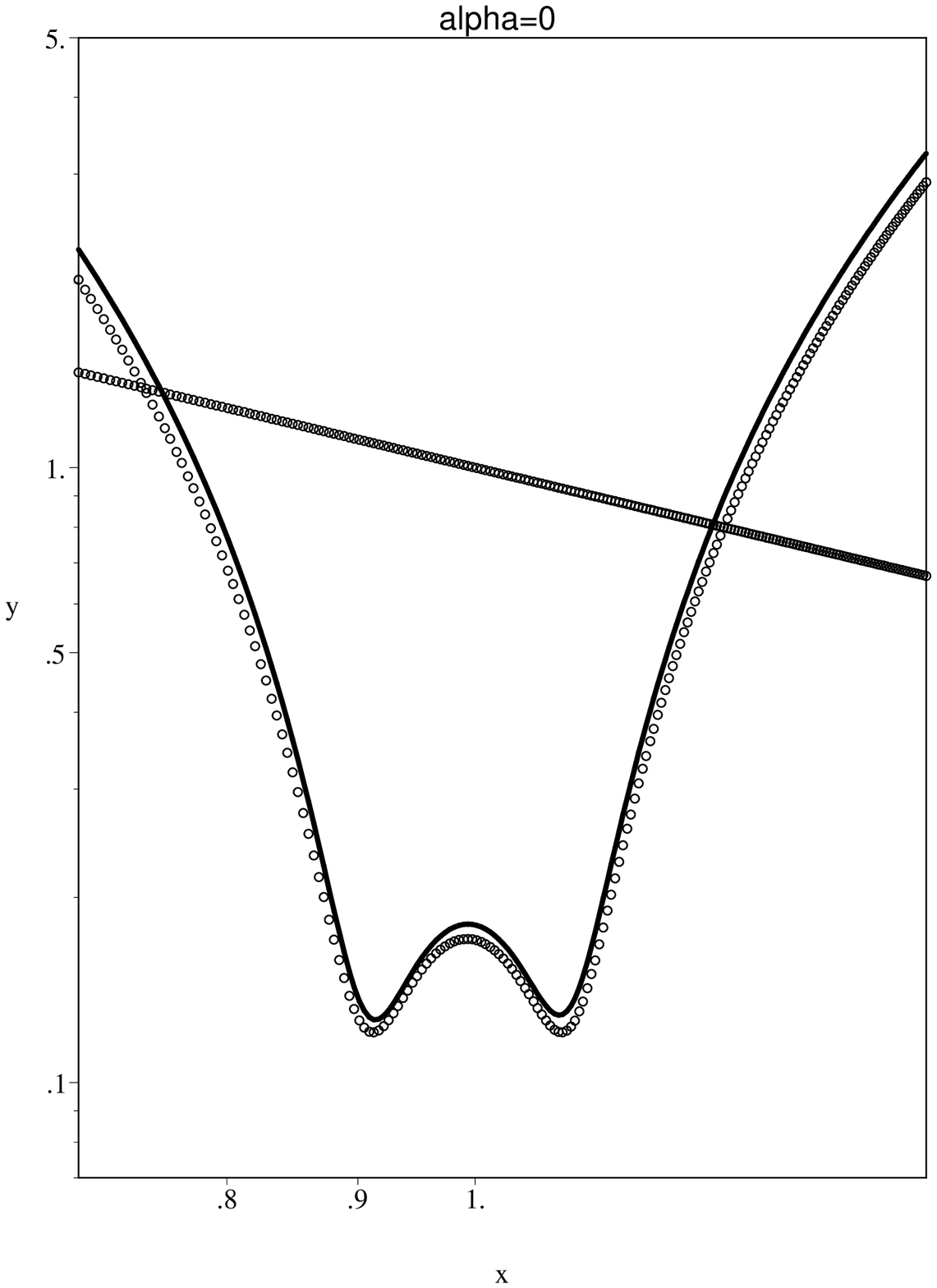}\quad
\includegraphics[width=0.23\textwidth,
    height=0.2\textwidth]{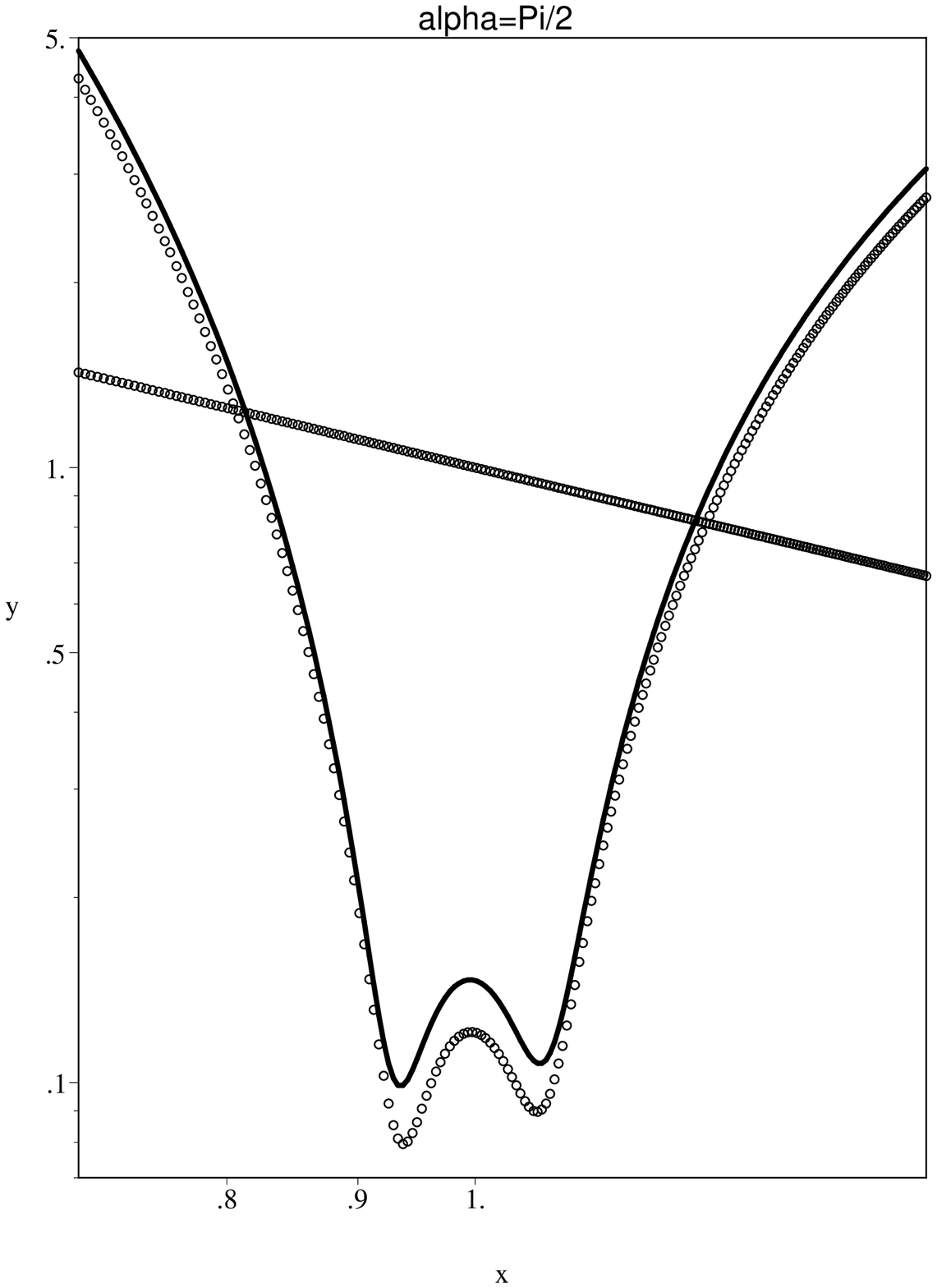}
\caption{Left plot:  sensitivity curves
$\xi$ plotted for parameters
$\alpha=0$, $\gamma_{0}/\Omega_0=.01$, $|{\cal A}_o|^2=0.24$ and
 $\eta=\eta_c$.
The dotted curves --- sensitivity
for no loss case with $\gamma_0^\text{no
 losses}=\gamma_{0}$.
Straight line depicts SQL sensitivity.
Right plot: the same as on the left
plot with homodyne angle $\alpha=\pi/2$. }\label{twoA}

\end{figure}

For $\eta >0$  we have two minima dependence of sensitivity $\xi$ at two
different frequencies $\Omega_\pm\simeq \Omega_0\sqrt{1\pm \eta}$.
With increase of $\eta$ the distance between
minima also increases. Comparing values $\xi(\Omega_\pm)$ with $\xi(\Omega_0)$
we can introduce a ``characteristic'' value $\eta_{\alpha\,c}$ when  $\sqrt
2\xi(\Omega_\pm) = \xi(\Omega_0)$:
\begin{equation}
  \eta_{\alpha\,c} = \frac{\xi_0}{C^{1/4}} \,.
\end{equation}
In this case
\begin{equation}
  \xi_0^2 = \frac{2\gamma_0^{\rm loss}}{\Omega_0|{\cal A}_\alpha|^2}\,C
  = \frac{\sqrt C}{2}\,\frac{\Delta\Omega^2}{\Omega_0^2} \,.
\end{equation}
[compare with Eq.\,(\ref{xi_enh_dbl})]. The sensitivity plots for this
case (again for  loss factor $|{\cal A}_0|^2=0.24$) are given in
Fig.\,\ref{twoA}.

\begin{figure}
\psfrag{.1}[rc][cl]{$0.1$}
\psfrag{.5}[rc][cl]{}
\psfrag{1.}[ct][cb]{$1$}
\psfrag{5.}[ct][lb]{$5$}
\psfrag{.1e2}{$10$}
\psfrag{.1e\2611}{$0.01$}
\psfrag{.4}[ct][cb]{$0.4$}
\psfrag{.6}[ct][cb]{$0.6$}
\psfrag{.7}[ct][cb]{$0.7$}
\psfrag{.8}[ct][cb]{$0.8$}
\psfrag{2.}[ct][cb]{$2$}
\psfrag{x}[lt][rb]{$\Omega/\Omega_{0}$}
\psfrag{y}[cc][lc]{$\xi$}
\psfrag{z}[cc][lc]{$\frac{Z(\Omega)}{m\Omega_{oo}^2}$}
\psfrag{2etac}[cb][ct]{$\alpha=0,\, \pi/2, \quad \eta=2\eta_c$}
\psfrag{alpha=Pi/2}[cb][ct]{$\alpha=\pi/2,\quad
 \eta=\eta_c,$}
\vspace{5mm}
\includegraphics[width=0.49\textwidth,
    height=0.35\textwidth]{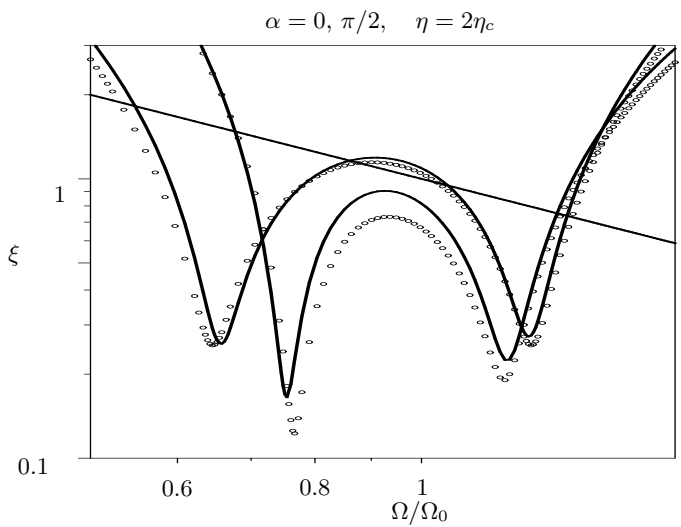}
\caption{Sensitivity curves $\xi$ plotted for parameters
 $\gamma_{0}/\Omega_0=0.03$, $|{\cal A}_o|^2=0.24$,  $\eta=2\eta_c$,
 $\alpha=0$ (curve with wider separated minimums) and $\alpha=\pi/2$.
The dotted curves --- sensitivity for no loss case with $\gamma_0^\text{no
 losses}=\gamma_{0}$.
Straight line depicts SQL sensitivity. }\label{wide}

\end{figure}

On Fig.~\ref{wide} we also present sensitivity curves for well separated
minimums ($\eta=2\eta_c$) and slightly less relaxation rate
$\gamma_0=0.03\Omega_0$ fot different values of homodyne angle $\alpha=0,\,
\pi/2$. It  is close to regime considered in \cite{Buonanno2001} including the
negligible role of optical losses (compare with plots on  Fig.~8 in
\cite{Buonanno2001}).

\subsection{Minimum of the spectral density}

Suppose that almost monochromatic signal has to be detected and we are
interested in the minimum of normalized spectral density $\xi$ at some fixed
frequency. First of all we have to remind that $\gamma_0^\text{loss}$ is fixed
(it depends on mirror's absorption only) whereas the effective loss factor $|{\cal
A}_0|^2$ (as well as $|{\cal A}_\alpha|^2$) can be modified by variation of
mirrors transmissivities.

It follows from Eq.\,(\ref{xi_real}) that $\xi(\Omega)$ reaches its minimum
at $\nu=\pm\eta_\alpha\Omega_0/2$ and $|{\cal A}_\alpha|=1$, and this
minimum is equal to
\begin{equation}
  \xi_{\rm min} = \sqrt{\frac{\gamma_0^{\rm loss}}{\Omega_0}
    \left(2+\frac{3}{\sqrt{2}}\right)}\simeq 0.06
\end{equation}
[compare with Eq.\,(\ref{xi_min_std})]. Here for the estimate we used
parameters from Table~\ref{tab1}.

Unfortunately, this value can not be obtained for planned
Advanced LIGO parameters. Indeed, it follows from
Eqs.\,(\ref{gamma_0load}),\,(\ref{delta_0}), that
\begin{equation}
  \delta_0<\frac{2\gamma_0^{\rm load}}{T_s^2} ,.
\end{equation}
(The physical sense of $\gamma_0^{\rm load}/T_s^2$ is quite
 clear ---
it is relaxation rate of a single FP cavity in one arm.)
In the double resonance regime we need to have
$\delta_0\approx\sqrt{2}\Omega_0\approx
10^3\ \text{s}^{-1}$. However for Advanced LIGO
 parameters
(Table~\ref{tab1})
\begin{equation}\label{large_gamma}
    \frac{2\gamma_0^\text{load}}{T_s^2}\simeq 100\
 \text{s}^{-1}
\end{equation}
It is less by one order of magnitude than required. The problem can be solved through modifying
Advanced LIGO parameters, namely, by decreasing the signal
recycling mirror transmittance $T_s^2$ by approximately one order of
magnitude and corresponding increase in the arm cavities input mirrors
transmittance by the same value.

\subsection{Signal-to-noise ratio}\label{sec:snr}

As was mentioned above, the double-resonance regime allows to obtain
signal-to-noise ratio better than SQL even for {\em wide-band} signals. It was
shown in \cite{05a1LaVy} that for no loss case the gain in signal-to-noise
ratio,  in principle, can be arbitrary high. Here we show that optical
losses restrict this gain. The details of calculations are presented  in
Appendix\,\ref{app:analysis:snr}.

As an example of wide band signal we consider the perturbation of metric
having the shape of a step function in time domain and the Fourier
transform equal to
\begin{equation}
\label{step}
h(\Omega)=\text{const}/\Omega.
\end{equation}
It is worth to underline that the result  practically
does not depend on the shape  of wide band signal spectrum (as alternative
example one could consider a short pulse (delta-function) --- its Fourier
transform is a constant).

To demonstrate the gain we also calculate the signal-to-noise ratio
$\text{SNR}_{\rm conv}$ for conventional LIGO interferometer
\cite{02a1KiLeMaThVy} without signal recycling mirror with registration of
phase quadrature and  take the quotient ${\cal P}$ of $\text{SNR}$ by
$\text{SNR}_{\rm conv}$ in order to characterize the gain in
signal-to-noise ratio (the value $\text{SNR}_{\rm conv}$ we calculated
numerically):
\begin{align}
{\cal P}&= \frac{\text{SNR} }{\text{SNR}_{\rm conv} },\quad
\text{SNR}_{\rm conv}\simeq 0.7\times
    \frac{h(\Omega_0)^2\Omega_0}{ h^2_{SQL}(\Omega_0)}
\end{align}

\begin{figure}
\psfrag{alpha=0}[cb][ct]{\hspace{-2cm}$\alpha=0$,\quad
    $\gamma_{0}/\Omega_0=0.0046$}
\psfrag{alpha=Pi/2}[cb][ct]{\hspace{-2cm}$\alpha=\pi/2$,\quad
    $\gamma_{0}/\Omega_0=0.0046$}
\psfrag{eta}{$\eta$}
\psfrag{y1}{$\cal P$}
\psfrag{0}[tc][bc]{$0$}
\psfrag{0.05}[tc][bc]{}
\psfrag{0.1}[tc][bc]{$0.1$}
\psfrag{0.15}[tc][bc]{}
\psfrag{0.2}[tc][bc]{$0.2$}
\psfrag{0.25}[tc][bc]{}
\psfrag{0.3}[tc][bc]{$0.3$}
\psfrag{0.35}[tc][bc]{}
\psfrag{0.4}[tc][bc]{$0.4$}
\psfrag{0.45}[tc][bc]{}
\psfrag{0.5}[tc][bc]{$0.5$}
\psfrag{2}[cc][lc]{$2$}
\psfrag{4}[cc][lc]{$4$}
\psfrag{5}[cc][lc]{$5$}
\psfrag{6}[cc][lc]{$6$}
\psfrag{7}[tc][bc]{}
\psfrag{8}[cc][lc]{$8$}
\psfrag{9}[tc][bc]{}
\psfrag{10}[cc][cc]{$10$}
\psfrag{11}[tc][bc]{}
\psfrag{15}[cc][cc]{$15$}
\psfrag{13}[tc][bc]{}
\psfrag{20}[cc][cc]{$20$}
\psfrag{25}[cc][cc]{$25$}
\psfrag{30}[cc][cc]{$30$}
\psfrag{35}[cc][cc]{}
\includegraphics[width=0.45\textwidth,
height=0.26\textwidth]{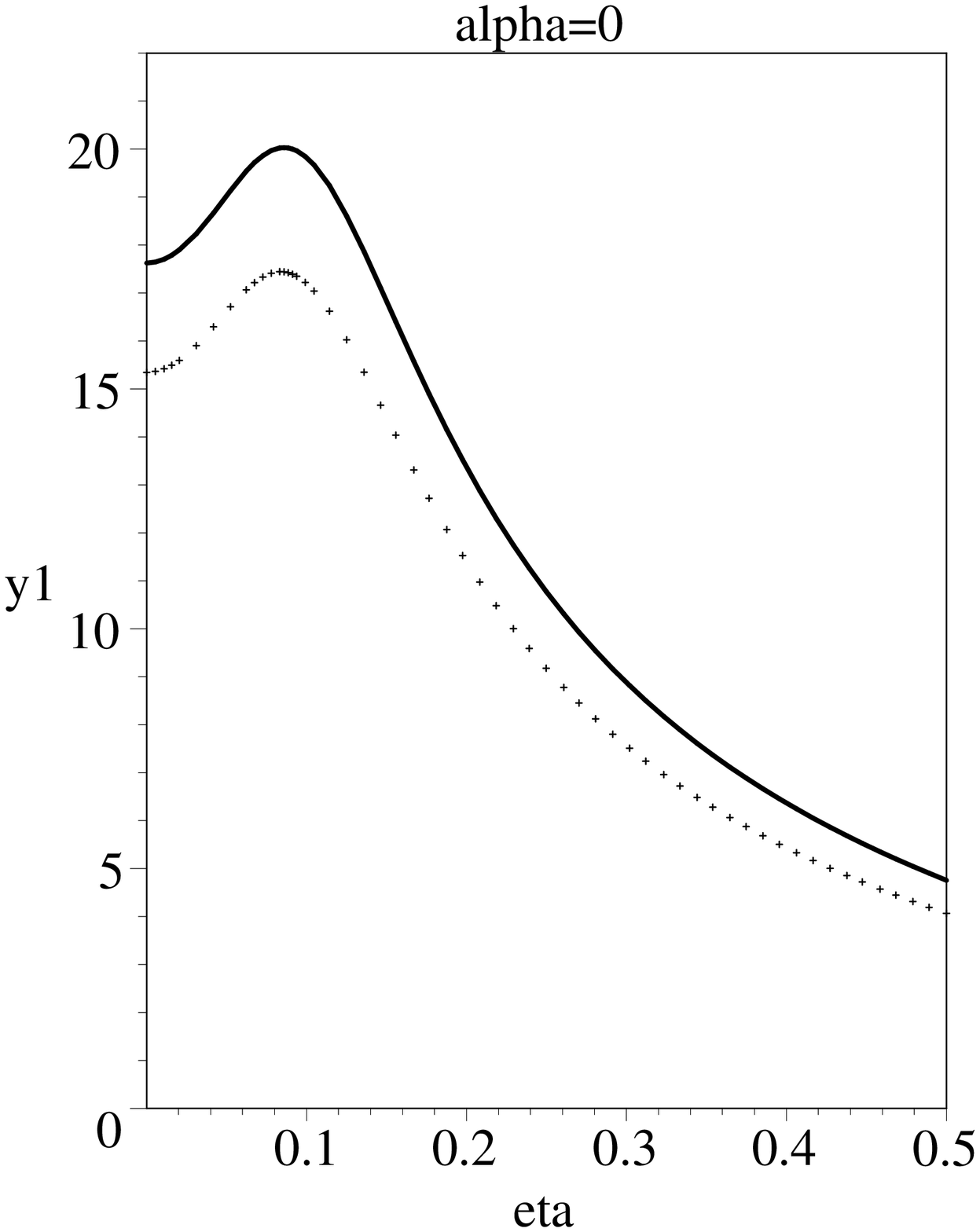}\\[8mm]
\includegraphics[width=0.45\textwidth,
height=0.26\textwidth]{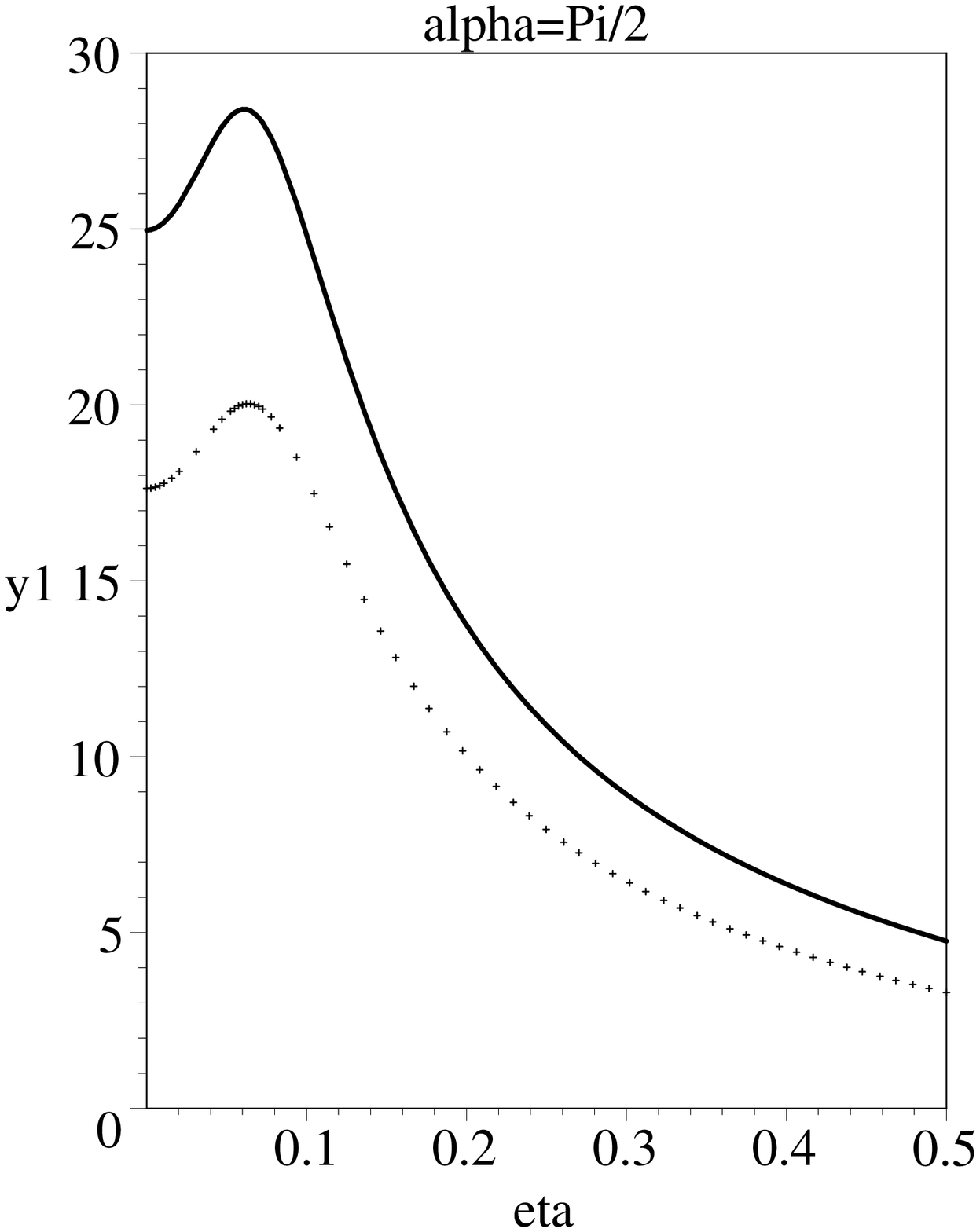}
\caption{Numerically calculated plots of signal-to-noise ratio  gain
${\cal P}(\eta)$ for different homodyne angles: $\alpha=0$ (top) and for
 $\alpha=\pi/2$ (bottom).
On each plot the upper solid curve corresponds to no loss
 case, the
lower dotted curve --- to planned in Advanced LIGO losses
 with factor
$|{\cal A}_0|^2=0.24$. All plots are calculated for fixed
 ratio
$\gamma_{0}/\Omega_0= 0.0046$, planned in Advanced LIGO.
 }\label{B2}
\end{figure}

Using the accurate formulas (\ref{xi})  we  numerically calculated plots (see
Fig.~\ref{B2}) for gain $\cal P$ as function of $\eta$ at fixed ratio
$\gamma_{0}/\Omega_0= 0.0046$  and loss factor factor $|{\cal A}_0|^2= 0.24$
corresponding to Advanced LIGO parameters (see Table~\ref{tab1}). We see that
the degradation due to losses is not large as compared with no loss case.

The gain $\cal P$ decreases with increase in $\eta$ bigger than optimal ---
i.e. when double-resonance ($\eta=0$) transforms to two well separated
first-order resonances ($\eta\gg \eta_c$).

We see from plots in Fig.~\ref{B2} that there is an optimal value of parameter
$\eta$ when gain has maximum. Analysis presented in Appendix
\ref{app:analysis:snr} shows that gain $\cal P$ reaches its maximum for optimal
values of $\eta_\alpha$ and $|{\cal A}|_\alpha$ and it is equal to
\begin{equation} \label{Pmax}
{\cal P}_\text{max}\simeq
0.6\times \sqrt \frac{\Omega_0}{\gamma_0^\text{loss}}\simeq 20
\end{equation}
where estimates given for parameters listed in Table~\ref{tab1}.

Note that this gain may be achieved at Advanced LIGO parameters --- in bottom
plot on Fig.~\ref{B2} the maximum of dotted curve is quite close to ${\cal
P}_\text{max}$.

It is worth noting that in the ``pure'' double-resonance
 regime ($\eta_\alpha=0$)
the gain in the signal-to-noise ratio only slightly
 differs from
the maximum gain:
\begin{equation}
\frac{{\cal P}_\text{max}}{{\cal P}(\eta_\alpha=0,\,
 |{\cal A}|_\alpha^\text{opt}) }=
        \frac{3^{3/4}}{2}\simeq 1.14
\end{equation}

Using the  signal-to-noise ratio for a conventional
oscillator (\ref{snr_oscill}), one can calculate the gain
\begin{equation}\label{Posc}
{\cal P}_\text{osc}=\frac{ \text{SNR}_\text{oscill} }{
        \text{SNR}_\text{conv}}\simeq 2.8
\end{equation}
Comparing (\ref{Pmax}) and (\ref{Posc}) we see that ``double resonance'' regime
provides gain in signal-to-noise ratio about $7$ times larger than conventional
oscillator.

\subsection{Squeezing in output wave}

\begin{figure}
\psfrag{r0=0}[cb][ct]{No losses: $|r_0|=0,\quad
 \eta=0,\quad \gamma_0^\text{no losses}/\Omega_0=0.01$}
\psfrag{r0=1}[cb][ct]{Optical losses: $|r_0|=1,\quad \eta=0,
 \quad
\gamma_{0+}/\Omega_0=0.01$}
\psfrag{x}[lt][rb]{$\Omega/\Omega_0$}
\psfrag{y}[cc][lc]{$\cal X$}
\vspace{7mm}
\includegraphics[width=0.46\textwidth,
    height=0.3\textwidth]{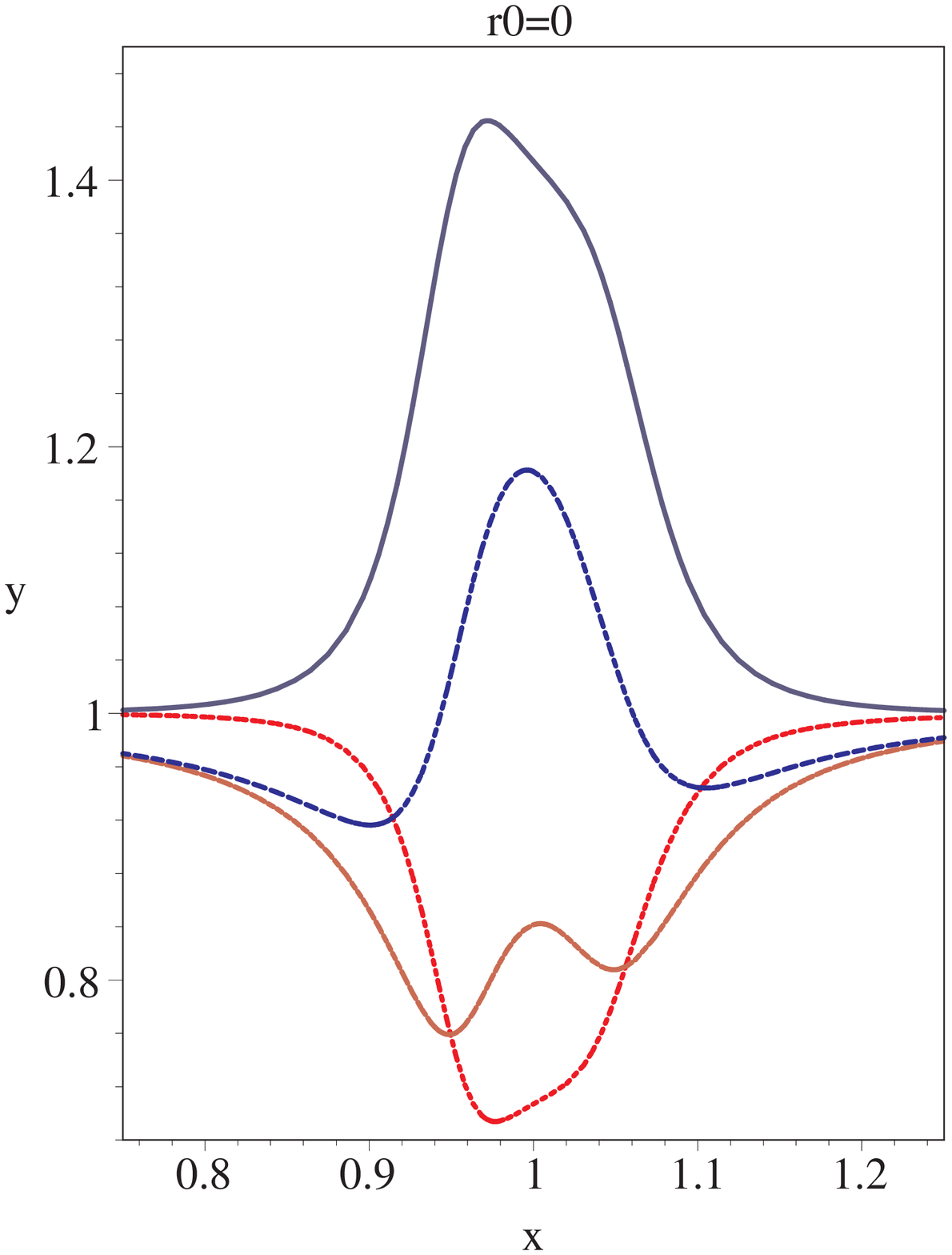}\\[8mm]
\includegraphics[width=0.46\textwidth,
    height=0.3\textwidth]{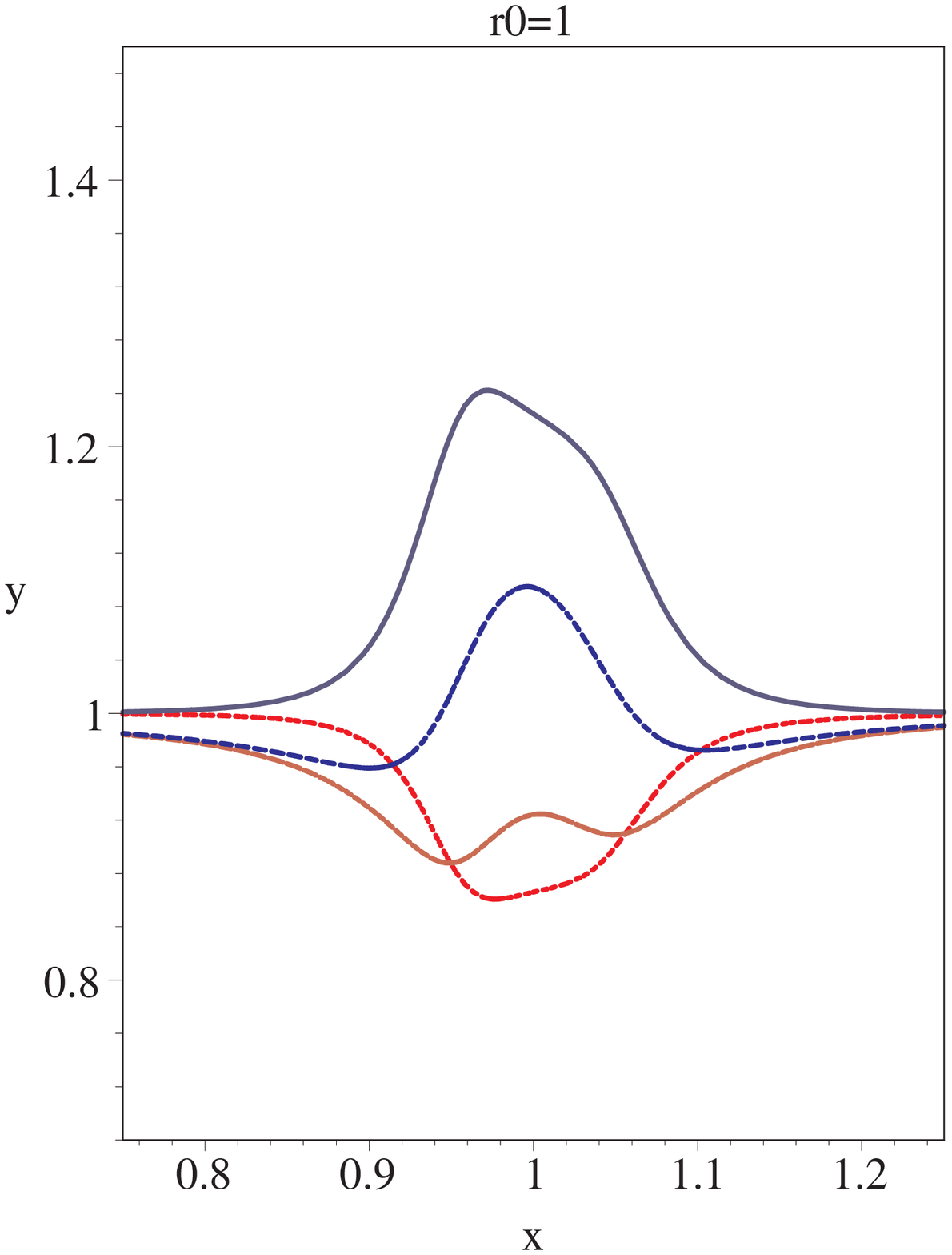}
\caption{The plots  of squeezing factor $\cal X$  as function on frequency
$\Omega$ for zero losses, homodyne angle varies from $\alpha =0$ (lower curve)
through $\alpha=\pi/6,\ \pi/3$ to $\alpha=\pi/2$ (upper curve) and  $\eta=0$.
Top: no loss case, $\gamma_0^\text{no losses}/\Omega_0=0.01$. Bottom: non-zero
losses case, $|r_0|=1$, $\gamma_{0+}/\Omega_0=0.01$. }\label{squ}
\end{figure}

It is also useful to find how optical losses affect
output wave squeezing. Using Eq.(\ref{bzeta}) one can calculate the
 squeezing factor
$\cal X$ (for the coherent quantum state, ${\cal X}=1$):
\begin{align}
{\cal X}(\Omega)&\equiv \sqrt{\frac{\langle
    b_\zeta b_\zeta^+ +     b_\zeta^+ b_\zeta
        \rangle}{\langle
        a_\text{vac} a_\text{vac}^+ + a_\text{vac}^+
 a_\text{vac}\rangle} }
\end{align}
The result of this calculation yields squeezing factor $\cal
 X$ as a
function of $\Omega$. For particular case $\eta=0$ the profile is
 given in
Fig.\,\ref{squ}. The top plot correspond to the lossless
 case (${\cal
A}_0=0$), the bottom one --- to the case when $|{\cal
 A}_0|=1$.

We see that in both cases the squeezing monotonously
 increases with
increase in homodyne angle. More important is the fact
 that the
values of squeezing factor are close to one. It confirms our
 assumption that
it is the optical rigidity rather than pondermotive
 nonlinearity, {\it
i.e.} the meter noises cross-correlation (as source of
 squeezing) that produces
the major input into the sensitivity gain.

\section{Conclusion}

The optical rigidity which can be created in the signal-recycled configuration
of laser interferometric gravitational-wave detectors turns the detector test
masses into oscillators and thus allows to obtain narrow-band sensitivity
better than the Standard Quantum Limit for a free test mass. This method of
circumnebting the Standard Quantum Limit does not rely on squeezed quantum
states of the optical filed and due to this it is much less vulnerable to
optical losses.

Moreover, sophisticated frequency dependence of this rigidity makes it
possible to implement the ``double resonance'' regime which provides
narrow-band sensitivity better than the Standard Quantum Limits for both a
free test mass and an conventional harmonic oscillator.

The ``double resonance''  regime may be useful to detect narrow band
gravitational waves, e.g. from pulsars.  Knowing pulsar parameters one can
tune the bandwidth and sensitivity in the  optimal way. It is important that
this tuning may be produced ``on line''  by varying signal recycling mirror
position and adjusting  circulating power.

Another advantage of the ``double resonance'' regime is its better
sensitivity to {\em wide-band} signals. While an conventional harmonic
oscillator provides approximately the same value of the the signal-to-noise
ratio as a free test mass, in the case of a ``double resonance'' oscillator
this parameter is limited only by the optical and mechanical losses and other
noise sources of non-quantum origin. Estimates based on the Advanced LIGO
parameters values shown that the ``double resonance'' regime can provide more
than tenfold increase of the wide-band signal-to-noise ratio.

\acknowledgments

We would like to extend our gratitude to  V.B.~Braginsky and Y.~Chen for
stimulating discussions. This work was supported by LIGO team from Caltech
and in part by NSF and Caltech grant PHY-0353775, as well as by the Russian
Foundation of Fundamental Research, grant No. 03-02-16975-a.

\appendix

\section{Analysis of Advanced LIGO interferometer.}\label{app:analysis}

\begin{figure}[t]

\psfrag{xE}[lc][lb]{$x_E$}
\psfrag{yE}[lc][lb]{$y_E$}
\psfrag{xN}[lc][lb]{$x_N$}
\psfrag{yN}[lc][lb]{$y_N$}

\psfrag{i}[cb][lb]{$i$}
\psfrag{-i}[lc][lb]{$-i$}

\psfrag{aP}[cb][lb]{$a_P$}
\psfrag{bP}[ct][lb]{$b_P$}
\psfrag{aW}[ct][lb]{$a_W$}
\psfrag{bW}[cb][lb]{$b_W$}

\psfrag{aD}[rc][lb]{$a_D$}
\psfrag{bD}[lc][lb]{$b_D$}
\psfrag{aS}[lc][lb]{$a_s$}
\psfrag{bS}[rc][lb]{$b_s$}

\psfrag{aE}[cb][lb]{$a_E$}
\psfrag{bE}[ct][lb]{$b_E$}
\psfrag{aE1}[cc][cc]{$a_{E1}$}
\psfrag{bE1}[cc][cc]{$b_{E1}$}
\psfrag{aE2}[cc][cc]{$a_{E2}$}
\psfrag{bE2}[ct][ct]{$b_{E2}$}

\psfrag{aN}[rc][lb]{$a_N$}
\psfrag{bN}[lc][lb]{$b_N$}
\psfrag{aN1}[lc][lc]{$a_{N1}$}
\psfrag{bN1}[rc][cc]{$b_{N1}$}
\psfrag{aN2}[rc][cc]{$a_{N2}$}
\psfrag{bN2}[lc][lt]{$b_{N2}$}
\psfrag{T}[lc][rc]{$T, \,r_1$}
\psfrag{r2}[cc][cc]{$r_2$}

\includegraphics[width=3.5in]{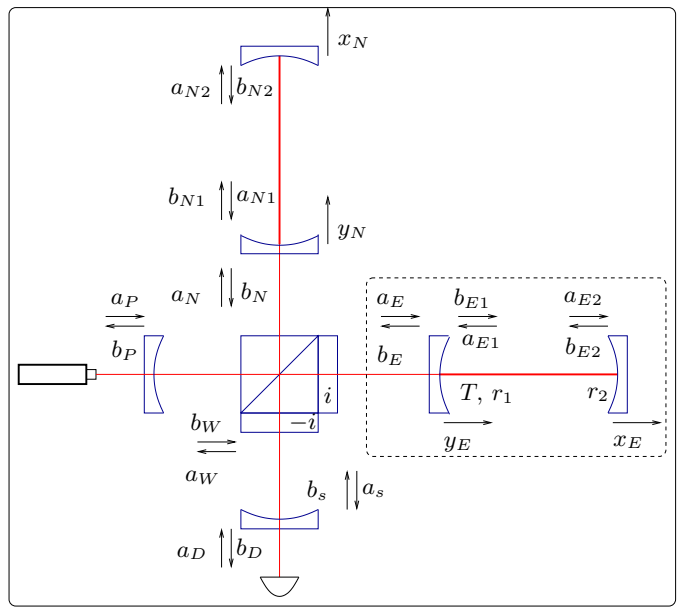}
\caption{Scheme of Advanced LIGO}\label{extraintra}
\end{figure}

\subsection{Notations and approximation}

We consider Advanced  LIGO interferometer with signal
 recycling (SR) mirror
and power recycling (PR) mirror as shown in
 Fig.\ref{extraintra}. SR, PR
mirrors and beam splitter are immobile and have no optical
 losses. We
assume that both Fabry-Perot (FP) cavities in the east and
 north arms are
identical: each input mirror has transmittance $T\ll 1$ and
 reflectivity
$R_1=\sqrt{1-T^2-{\cal A}_1^2}$, each end mirror has
 reflectivity
$R_2=\sqrt{1-{\cal A}_2^2}$, where ${\cal A}_1,\,{\cal A}_2$
 are optical
loss coefficients for input and end mirrors
 correspondingly. The  end
mirrors and input mirrors have equal masses $m$ and they can
 move as free
masses.

The contribution of losses in SR mirror and in beam splitter
 is small as
compared with contribution of  losses in mirrors in arms
 which is larger by
factor $\sim 1/T^2$ --- consequently we assume in our
 analysis that SR mirror
and beam splitter have no optical losses.

We consider ``dark port'' regime when no regular optical
 power comes to
detector in the signal port. In this case the power recycling
 mirror is used only
to increase the mean power delivered to the beam splitter
 and no fluctuational
fields from the laser (west arm) access detector in the
 south arm.

Upper letters denote  the mean complex amplitudes, lower
 case  letters
denote operator describing  fluctuations and signal.
 south arm, $b_D$ is the operator of
 Fig.\ref{extraintra}).
The
electric field $E$ in propagating wave
can be written as sum of large mean field and small component (see
details in \cite{02a1KiLeMaThVy}):

\begin{align*}
E & \simeq  \sqrt{\frac{2\pi \, \hbar \omega_o}{Sc}
 }\,e^{-i\omega_o t}
        \left( A +\int_{-\infty}^\infty
         a\,e^{-i\Omega t} \frac{d\Omega}{2\pi}\right)+
        \{\mbox{h.c.} \},\\
I & =  \hbar \omega_p|A|^2,\quad
    a\equiv a(\omega_o+\Omega),\quad a_-\equiv
 a(\omega_o-\Omega)\\
&        \big[a(\omega_o\pm\Omega),\,a^+(\omega_o\pm
 \Omega')\big]=
    2\pi\,\delta_0(\Omega -\Omega'),
\end{align*}
where $A$ is the complex amplitude, $S$ is the cross section of
 the light
beam, $c$ is the velocity of light, $I$ is
 the mean power of the light
beam, $a$ and $a^+$ are the annihilation and creation operators.
  We
consider sidebands $\omega_o\pm \Omega$ about carrier
 $\omega_o$ with
side band frequencies $\Omega$ in gravitational wave range
($\Omega/2\pi \in 10\dots 1000$~Hz). Detailed notations of
 wave
amplitudes and mirror displacements are given on
 Fig.\ref{extraintra}.

\subsection{Arm cavities}

First we consider the fields propagating in FP cavity in
 the east arm shown  in
dashed box on Fig.\ref{extraintra} (formulas for FP cavity
 in the north arm are
the same with obvious substitutions in subscripts $E\to N$).
 Below we assume that
the following conditions are fulfilled:
\begin{align}
\label{smallt}
\frac{L\Omega}{c}\ll 1,\quad  T,\ {\cal A}_1, \ {\cal
 A}_2\ll 1,
\end{align}
where $L$ is the distance between the mirrors in arms
 ($4$~km for LIGO).

We start with a set of equations for mean amplitudes
\begin{align}
B_{E1}&= iT A - R_1 A_{E1},\quad
B= iT A_{E1} - R_1 A_E,\\
A_{E1}&=  B_{E2}\, e^{i\omega_o L/c} =-R_2 B_{E1}\,
 e^{2i\omega_o L/c}.
\nonumber
\end{align}
Below we assume that the carrier frequency of the incident
 wave  $\omega_o$ is equal to  the
eigenfrequency of each FP cavity, i.e. $e^{i\omega_o
 L/c}=1$. So we obtain
\begin{align}
B_{E1}& = {\cal T} A_E,\quad B_{E2}\simeq -B_{E1}, \quad
 B_E= {\cal R} A_E,\\
{\cal T} &\equiv\frac{iT}{1-R_1R_2}\simeq\frac{2i\gamma_{\rm
 load}}{T\gamma},\\
{\cal R}&\equiv
        \frac{-R_1+R_2(1-r_1^2)}{1-R_1R_2}\simeq
        \frac{\gamma_-}{\gamma},\\
\gamma &= \gamma_{\rm load} + \gamma_{\rm loss}, \quad
\gamma_- = \gamma_{\rm load} - \gamma_{\rm loss}, \nonumber
 \\
\gamma_{\rm load} &= \frac{cT^2}{4L}, \quad
\gamma_{\rm loss} = \frac{c{\cal A}^2}{4L} \,, \nonumber \\
{\cal A} &= \sqrt{{\cal A}_1^2+{\cal A}_2^2} \,. \nonumber
\end{align}

To calculate fluctuational components of field we start from
 the set of
equations:
\begin{align}
\label{fl2a}
b_{E1}&= iT a_E - R_1 a_{E1} +i{\cal A}_1 e_{E1}
 +R_1A_{E1}\,2iky_E,\\
\label{fl2b}
b_E&= iT a_{E1} - R_1 a_E  -R_1A_E\, 2iky_E +i{\cal A}_1
 e_{E1},\\
\label{fl2c}
a_{E1}&=\theta R_2\left(
        -a_{2E}\,  -A_{2E}\,2ikx_E\right) +i{\cal
 A}_2e_{2E}.
\end{align}
Here  $e_{1E,}\ e_{2E}$ is vacuum fields appearing due to
 losses.
We denote $\theta=e^{i(\omega_o+\Omega) L/c}$   using
obvious approximation $\theta \simeq 1+i\Omega L/c$,

Substituting (\ref{fl2c}) into (\ref{fl2a}) and using
 conditions
(\ref{smallt}) we obtain:
\begin{align}
\label{b2rr}
 b_{E1}&\simeq {\cal T}_{\Omega}\, a_E+ e_E r\,{
 \cal T}_\Omega+
    \frac{{\cal T}_{\Omega}\,B_{E1}\, 2ik(x_E-y_E)}{iT} ,\\
b_{2E}&\simeq -b_{E1}, \quad a_{2E}\simeq b_{E1},\quad
 a_{E1}\simeq - b_{E1},\\
{\cal T}_{\Omega} &\equiv\frac{iT}{1-R_1R_2\theta^2}\simeq
        \frac{2i\gamma_{\rm
 load}}{T\big(\gamma-i\Omega\big)},\\
e_E&=\frac{e_{E1}{\cal A}_1- e_{2E}{\cal A}_2}{{\cal A}},
    \nonumber
\end{align}
Here we introduced  operator $e_E$ of vacuum fluctuations,
 which fulfill usual
commutator relation $\big[e_E(\omega_o+\Omega), \,
e_E^+(\omega_o+\Omega')\big]=2\pi\, \delta_0(\Omega-\Omega')$.

Substitution of (\ref{b2rr}) into (\ref{fl2b}) under
 conditions
(\ref{smallt}) allows us to get the formula for reflected field:
\begin{align}
\label{2Fe}
b_E &= a_E\,{\cal R}_{\Omega} - ierT\,{\cal T}_\Omega -
 A{\cal T}{\cal
        T}_\Omega \, 2ik(x_e-y_E),\\
{\cal R}_{\Omega} &=
        \frac{-R_1+R_2(1-{\cal
 A}_1^2)\theta^2}{1-R_1R_2\theta^2}\simeq
        \frac{\gamma_-+i\Omega}{\gamma-i\Omega}
\end{align}

\subsection{Beam splitter}

We assume that lossless beam splitter has transmittance and
 reflection factors
equal to $i/\sqrt2$ and $-1/\sqrt2$, correspondingly. The
 additional phase
shift $\pi/2$ is added to the west arm and $-\pi/2$ to the
 south one.
It is convenient to introduce new variables
\begin{equation}
  a_{(\pm)} = \frac{a_E\pm a_N}{\sqrt2} \,, \qquad
  b_{(\pm)} = \frac{b_E\pm b_N}{\sqrt2} \,.
\end{equation}
For these variables beam splitter equations read:
\begin{subequations}\label{eil_bs}
  \begin{gather}
    a_W = -b_{(+)} \,, \qquad
    a_s = -b_{(-)} \,, \\
    a_{(+)} = -b_W \,, \qquad
    a_{(-)} = -b_s \,.
  \end{gather}
\end{subequations}
These equations are valid both for the zero and the first
approximations as the beam splitter is a linear system.

Now we can consider the fields in entire interferometer.
For our mean amplitudes of $(\pm)$ variables we have
  \begin{align}
 B_{(\pm)}&={\cal R} A_{(\pm)}
  \end{align}
 It is easy to note that the set of equations  splits into
 two
independent sets: one for $W$ and $(+)$ and second for $S$
 and $(-)$. The
``$S,(-)$'' set has only trivial zero solution since
 $A_D=0$. Therefore,
{\allowdisplaybreaks\begin{subequations}\label{eil_0_symm}
  \begin{align}
     A_E &= A_N = \frac{A}{\sqrt2} \,, \quad
     B_E = B_N = \frac{{\cal R} A}{\sqrt2} \,, \\
     A_{E1} &= A_{N1} =
    \frac{-\Theta^2B_1}{\sqrt 2} \,, \quad
     B_{E1} = B_{N1} = \frac{B_1}{\sqrt2}\,,
  \end{align}
\end{subequations}}
>From this point on we omit  indices $+$ for the sake of simplicity, $A$ is
 the complex mean
amplitude left in the beam splitter.

\subsection{Output field}

The signal wave, registered by detector in the south arm (``$S,\
 (-)$'' mode) is
coupled with the differential motion of the mirrors, i.e. it
 has a part
proportional to differential coordinate
\begin{equation}
  x=\frac{(x_E-y_N)-(x_N-y_N)}{2}
\end{equation}
coupled with the gravitation-wave signal. We are interested
 in ``$S,\,
(-)$'' mode and below we omit subscripts $_{(-)}$ for
 $a_{(-)},\ b_{(-)} $.

>From equations for the south arm
\begin{subequations}
\label{eil_s_1s}
  \begin{gather}
    b_D = -R_s a_D + iT_s\theta_s a_s \,, \\
    b_s = -R_s\theta_s^2 a_s + iT_s\theta_s a_D \,.
\end{gather}
\end{subequations}
(here $\theta_s=e^{i(\omega_o+\Omega)l/c}$, $l$ is the optical
 length between SR
mirror and input mirror in the arm) one can obtain the following
 formulas for
fluctuational component of the output field:
\begin{align}\label{bD1}
  b_D &\simeq {\cal R}_s\,a_D - {\cal T}_s{\cal T}_\Omega
    (2ikB_1x + i{\cal A}\,e),\\
  e & =\frac{e_E-e_N}{\sqrt 2}
\end{align}
Here we introduced the "generalized" transparency ${\cal
 T}_s$ and
reflectivity ${\cal R}_s$ of "$S,\ (-)$" mode of
 interferometer:
\begin{align}
\label{eil_1s_4}
{\cal T}_s &= \frac{-iT_s\theta_s}{ 1+{\cal R}_\Omega
 R_s\theta_s^2},\quad
{\cal R}_s = \frac{-\big(R_s +{\cal
 R}_\Omega\theta_s^2\big)}{ 1+
        {\cal R}_\Omega R_s\theta_s^2},
\end{align}
 Due to small value of $l\ll L$ we
consider below $\theta_s$ a constant not depending on
 $\Omega$.

It is useful to  rewrite formulas (\ref{eil_1s_4}) in
 approximation
(\ref{smallt}) denoting  $\theta_s=i\,e^{i\phi}$:
\begin{align}
{\cal R}_s &\simeq
 G\times\frac{\Gamma_-^*+i\Omega}{\Gamma-i\Omega}, \quad
  G=\frac{R_s+e^{2i\phi}}{1+R_s e^{2i\phi}}\, \\
\Gamma &= \gamma_0-i\delta_0,\quad
  \Gamma_- = \gamma_{0-}-i\delta_0, \\
\label{gamma0p}
\gamma_0 &= \gamma_0^{\rm load} + \gamma_0^{\rm loss}, \quad
  \gamma_{0-}=\gamma_0^{\rm load} - \gamma_0^{\rm loss}, \\
\label{gamma_0load}
\gamma_0^{\rm load}
  &= \frac{\gamma_{\rm load}\,T_s^2}{\big|1+R_s e^{
 2i\phi}\big|^2}
  = \frac{c\,T^2T_s^2}{4L\big|1+R_s e^{ 2i\phi}\big|^2}\,,
  \\
\label{gamma_0loss}
\gamma_0^{\rm loss} &= \gamma_{\rm loss}=\frac{c{\cal
 A}^2}{4L}, \\
\label{delta_0}
\delta_{0} &= \frac{2R_s\gamma_{\rm load}\sin 2\phi}{
        \big|1+R_s e^{2i\phi}\big|^2}=
        \frac{2R_s\gamma_0^{\rm load}\sin 2\phi}{
        T_s^2}\\
{\cal T}_s &= \frac{-iT_s\theta_s}{ 1+{\cal R}_\Omega
 R_s\theta_s^2}=
    \frac{T_s\,e^{i\phi}
        \big[\gamma-i\Omega\big]}{
 \big[1+R_s\,e^{2i\phi}\big]\big[\Gamma-i\delta_0\big]}\,
        .\nonumber
\end{align}
Here $\delta_0$ is the detuning, $\gamma_0$ is the relaxation rate
 of our
difference mode (``$S,\,(-) $'' mode) which can be presented
 as a sum of
``loaded'' and ``intrinsic'' rates.  ``Intrinsic''
 relaxation rate
$\gamma_0^{\rm loss}$ is exclusively provided by intrinsic
 losses in
mirrors in arms while ``loaded'' relaxation rate
 $\gamma_0^{\rm load}$ is
provided only by transparencies of SR and input mirrors.
 Note that for
Advanced LIGO relaxation rate $\gamma_0\simeq 2\
 \text{s}^{-1}$ --- it is
much less than the mean frequency of gravitational wave range
 $\Omega \sim
2\pi\times 100\ \text{s}^{-1}$.

\subsection{Pondermotive forces}

To calculate pondermotive force acting on movable
 mirrors in interferometer arms
we should write down
the equation for differential field
$b_1=\big(b_{E1}-b_{N1})/\sqrt 2$\,   in approximation
 (\ref{smallt}):
\begin{align}
b_1&\simeq {\cal T}_s{\cal T}_\Omega\left(
        a_D  +\frac{\big(1+R_se^{2i\phi}\big) }{T_s
 e^{i\phi}}
        \left[re+\frac{2B_1kz}{T}\right]\right)
\end{align}

The incident wave  acts  on the mirror with force
proportional to square of amplitude module; and we keep only
 the cross term
of this square.
\begin{align}
{\cal F}&\simeq \int_{-\infty}^\infty  F(\Omega)
        \, e^{-i\Omega t}\, \frac{d\Omega}{2\pi},
        \quad
F(\Omega)=\hbar k\,\big(A^*a +A a^+_-\big)  \nonumber \
\end{align}

First we write down the formulas for forces acting on the back
 mirrors.
The difference between the forces acting on the east and
north back mirrors is  equal to:
\begin{align}
 F_2 & \simeq 2\hbar k\,\big( A^*_{2}a_{2}+A_{2}a^+_{2-}+
        B^*_{2}b_{2}+B_{2}b^+_{2-}\big),
\end{align}
where $A_2=(A_{E2}-A_{N2})/\sqrt 2, \
 a_2=(a_{E2}-a_{N2})/\sqrt 2 $
and so on (recall that we continue considering "$S\, (-)$"
 mode).
One can extract two terms in formula for $F_2$:
\begin{equation}
  F_2 = F_{\rm meter} + F_{\rm rigid} \,,
\end{equation}
where the first term corresponds to fluctuational component
 (back action)
and the second one --- to the regular force depending on mirrors
 positions
(optical rigidity). The formula for $F_{\rm meter}$ is the
 following
\begin{multline}\label{faD2a}
  F_{\rm meter} \simeq \frac{2i\hbar\omega_o T_sT\,\,B_1^*
 \, e^{i\phi}
    \big(a_{D} +{\cal A}_0e\big)}{
      \big(1+R_s\,e^{2i\phi}\big)\big(\Gamma-i\Omega\big)} \\
    - \frac{2i\hbar\omega_o T_sT\,\,B_1\,e^{-i\phi}
        \big(a_{D-}^+ +{\cal A}_0^*e^+_-\big)}{
    \big(1+R_s\,e^{-2i\phi}\big)\big(\Gamma^*-i\Omega\big)} \,,
\end{multline}
where
\begin{equation}\label{r0}
  {\cal A}_0 = \frac{{\cal
 A}\big(1+R_s\,e^{2i\phi}\big)}{TT_se^{i\phi}}
\end{equation}
is the effective loss factor.

The equation for optical rigidity $K$ has the following shape
\begin{align}
\label{K}
K&\equiv \frac{-F_\text{rigid}}{x}=
    \frac{16\hbar k^2 \, \gamma_{\rm
 load}\delta_{0}|B_1|^2}{T^2\cal D}=
        \frac{8k\,I_c \delta_{0}
        }{L{\cal D}},\\
{\cal D}&= (\Gamma-i\Omega)(\Gamma^*-i\Omega), \quad
    I_c =\frac{\hbar \omega_o}{2}|B_1|^2,
\end{align}
We see that optical rigidity  $K$ is proportional to
 detuning $\delta_0$ of
the difference mode of the interferometer. This detuning can be
 only introduced
by displacement of SR mirror (while Fabry-Perot
 cavities in arms
remain in optical resonance).

In  approximation (\ref{smallt}) the forces acting on the back
 mirrors are
approximately equal (with negative sign) to the forces acting on
 input mirrors
--- the difference is negligible. So for the difference coordinate
 $x$ we have
the following equation:
\begin{equation}\label{zZa}
  Z(\Omega)x = F_{\rm meter} + F_{\rm signal} \,,
\end{equation}
where
\begin{equation}
  Z(\Omega) = -m\Omega^2 + K
\end{equation}
and
\begin{equation}
  F_{\rm signal} = \frac{m\Omega^2 Lh}{2}
\end{equation}
is the signal force due to action of gravitational wave, $h$  --- is
the dimensionless gravitational-wave signal.

\subsection{Output signal}

We see from formula (\ref{faD2a}) that the back action
force is produced by sum of  fluctuational fields: $a_D$
 from
signal port and $e$ due to losses in mirrors.
It is useful to introduce a new pair of independent
fluctuational fields $p$ and $q$ (the new basis) as following:
\begin{subequations}\label{ae_to_pq}
  \begin{align}
    p &= G\,\frac{\Gamma^*+i\Omega}{\Gamma-i\Omega}
      \frac{a_D + {\cal A}_0e}{\sqrt{1+|{\cal A}_0|^2}} \,, \\
    q &= \frac{-{\cal A}_0^*a_D+e}{\sqrt{1+|{\cal A}_0|^2}} \,.
  \end{align}
\end{subequations}
These two bases are equivalent --- both pairs $a_D,\ e$ and
 $p,\ q$
describe vacuum fluctuations (we do not consider possible
 squeezing
of field $a_D$).

Then we can rewrite formulas for the fluctuational force
 $F_{fl}$ and output
field $b_D$ in a more compact form:
\begin{subequations}
  \begin{align}
    F_{fl} & \simeq\frac{8i\hbar kB_1^*\gamma_0
      \big(1+R_se^{2i\phi}\big)}{TT_s\, e^{i\phi}
        \big(\Gamma^*+i\Omega\big)}\times
          \frac{p}{\sqrt{1+|{\cal A}_0|^2}}+
            \big\{\mbox{h.c.}\big\}_{-\Omega} \,, \\
    b_D &= \frac{p}{ \sqrt{1+|{\cal A}_0|^2}}+
        \frac{q\, {\cal A}_0^*}{\sqrt{1+|{\cal A}_0|^2}}-
        {\cal T}_s {\cal T}_\Omega B_1 ikz. \label{bDa}
  \end{align}
\end{subequations}

Calculating the difference position $x$ from (\ref{zZa}) and
 substituting it
into (\ref{bDa}) one can obtain
the final formula for the output field $b_D$:
\begin{align}
  \label{bD2a}
 b_D& = \frac{q\, {\cal A}_0^*}{\sqrt{1+|{\cal A}_0|^2}} +
      \frac{-m\Omega^2}{Z(\Omega)\sqrt{1+|{\cal
 A}_0|^2}}\times\\
&      \times\left\{
        p\left(
          1-\frac{iJ\Gamma^*}{\Omega^2\big((\Gamma^*)^2
 +\Omega^2\big)}
       \right)\right.+\nonumber\\
&\quad \left.      + p_-^+\left(
          \frac{-iJ\gamma_0G}{\Omega^2\big(\Gamma^2
        +\Omega^2\big)}\right)
          +
 \frac{ih\sqrt{2J\gamma_0G}}{\Omega\,h_{SQL}\big(\Gamma-i\Omega\big)}
              \right\} \,,\nonumber
\end{align}
where
\begin{equation}
  J = \frac{8kI_c}{mL} \,.
\end{equation}

This formula has two terms. The first one ($\sim q$) is
 proportional to
the effective loss factor ${\cal A}_0$ and appears due to
 the optical
losses. The second term  has the same form as for no losses
 case ---
compare with Eq.(16) in \cite{05a1LaVy} --- with weighted
 multiplier
$1/\sqrt{1+|{\cal A}_0|^2}$ and substituted damping rate
 $\gamma_0$
accounting for losses in \cite{05a1LaVy}. It is the background to
 interpret formula
(\ref{bD2a}) as field reflected from lossless interferometer
 (with
damping rate $\gamma_{0}$) and then passed through grey
 filter with
total loss factor $|{\cal A}_0|$ which decreases our field
 and adds
fluctuations (the first term).

It is important that the second term has  multiplier $
 m\Omega^2/Z(\Omega)$ which
increases the relative contribution of the second term at
 frequencies close to mechanical
resonance (when $|Z(\Omega)|\ll  m\Omega^2$) while the first
 term does not depend
on frequency --- it demonstrates the advantage of ``real''
 rigidity and
may explain the relatively weak
degradation of sensitivity due to optical losses.

\subsection{Sensitivity}

In gravitational wave antenna one registers the quadrature
 component
$b_\zeta$ of output fields using the balanced homodyne scheme
 (not
shown in Fig.\ref{extraintra}):
  \begin{subequations}
  \label{bzeta}
    \begin{align}
      b_\zeta
 &=\frac{b_De^{-i\zeta}+b_{D-}^+e^{i\zeta}}{\sqrt2}
      \nonumber\\
       & = \frac{-m\Omega^2}{(Z(\Omega)\sqrt {2(1+|{\cal
 A}_0|^2}}
          \left\{
            A_q\left(q{\cal A}_0^*e^{-i\zeta} +
                q_-^+{\cal
 A}_0e^{i\zeta}\right)\right.\nonumber\\
       &\qquad \left.     + A_p\,p +A_p^*\,p^+_-
            + A_s\frac{h_s}{h_{SQL}(\Omega_0)}
          \right\}\,, \label{bzeta1} \\
      A_q &= \frac{-m\Omega^2+K}{-m\Omega^2}
        = 1-\frac{J\delta_0}{\Omega^2{\cal D}}\,,\\
      A_p &= \frac{e^{-i\zeta\left[
          (\Gamma^*_+)^2+\Omega^2 -iY\Gamma^*_+ +
 iY\gamma_{0+}e^{2i\alpha}
        \right]}}{(\Gamma^*_+)^2+\Omega^2}\,, \\
      A_s &= \frac{\sqrt{J\gamma_0G}}{\Omega_0}
         \left(\frac{ie^{-i\alpha}}{\Gamma_+-i\Omega}-
           \frac{ie^{i\alpha}}{\Gamma_+^*-i\Omega}\right)=
           \nonumber\\
       &= \frac{\sqrt{8J\gamma_0}}{\Omega_0}
 \frac{(\gamma_0-i\Omega)\sin\alpha-\delta_0\cos\alpha}{{\cal D}}
       \,,\\
      \alpha &= \zeta-\phi \,.
    \end{align}
  \end{subequations}
Here and below we normalize dimensionless metric $h_s$
 by SQL sensitivity
$h_{SQL}(\Omega_0)$ at some frequency $\Omega_0$.

Now we can write down one-sided spectral noise
density $S_h$ recalculated
 to variation of dimensionless metric $h_s$ and
 normalize it to
SQL sensitivity $h_{SQL}(\Omega_0)$:
\begin{subequations}\label{xi}
\begin{align}
\xi^2(\Omega)&= \frac{S_h(\Omega)}{h_{SQL}(\Omega_0)^2}=
    \frac{2|A_q|^2+2|A_p|^2}{|A_s|^2}=\frac{P_1+P_2+P_3}{Q}, \\
P_1&= \Big[ \Omega^4 -\Omega^2(\delta_0^2-\gamma_{0+}^2)
  + J\big(\delta_0 - \gamma_0\sin 2\alpha\big)\Big]^2 \,, \\
P_2&=\gamma_0^2\big(2\delta_0\Omega^2 - J(1- \cos
 2\alpha)\big)^2,\\
P_3&= |{\cal
 A}_0|^2\Big\{\big[\Omega^4-(\delta_0^2+\gamma_0^2)\Omega^2
 +
        J\delta_0\big]^2 + 4\gamma_0^2\Omega^6\Big\}, \\
Q&=\frac{4J\gamma_0\Omega^4}{\Omega_0^2}
        \big|(\gamma_0-i\Omega)\sin\alpha -
    \delta_0\cos\alpha\big|^2,
\end{align}
\end{subequations}
Note that without signal
recycling mirror SQL sensitivity in LIGO lossless interferometer
 can be achieved at working frequency
 $\Omega_0=\gamma$ if the
optical power $I_c$ is equal to the optimal one
 $I_{SQL}(\Omega_0)$
\cite{02a1KiLeMaThVy,Buonanno2001}:
\begin{equation}
\label{ISQL}
I_{SQL}(\Omega_0)=\frac{m\Omega_0^3 Lc}{8\omega_o},\quad
 \text{or}\
    \frac{J}{\Omega_0^3}=1.
\end{equation}

Although presentation (\ref{xi}) is compact and convenient for
 numeric
estimates, it can mask the physical structure of the noise. Due to
 this reason,
we provide a more transparent form of this equation:
\begin{equation}
  S_h(\Omega) = \frac{4}{m^2L^2\Omega^4}\left[
    \frac{\hbar^2}{S_x} + |Z_{\rm eff}|^2S_x + |Z|^2S_{\rm
 loss}
  \right] \,,
\end{equation}
where
\begin{equation}
  S_x = \frac{\hbar}{2mJ\gamma_0}\frac{|{\cal D}|^2}
    {\bigl|(\gamma_0-i\Omega)\sin\alpha -
    \delta_0\cos\alpha\bigr|^2}
\end{equation}
is the measurement noise,
\begin{equation}
  S_{\rm loss} = |{\cal A}_0|^2S_x
\end{equation}
is the noise created by the optical losses,
\begin{equation}
  Z_{\rm eff} = K_{\rm eff} - m\Omega^2 \,,
\end{equation}
and
\begin{multline}\label{K_eff}
  K_{\rm eff} = \frac{m J}{|{\cal D}|^2}\,\Bigl[
    \delta_0(\gamma_0^2+\delta_0^2-\Omega^2) \\
    + \gamma_0(\gamma_0^2-\delta_0^2+\Omega^2)\sin2\alpha
    - 2 \delta_0\gamma_0^2\cos2\alpha
  \Bigr] \,.
\end{multline}
is the effective rigidity.

\subsection{Narrow-band case}

Suppose that the observation frequency is close to the  double resonance
frequency $\Omega_0=\delta_0/\sqrt{2}$ and the pumping power is close to
the critical value:
\begin{subequations}
  \begin{gather}\label{Jinit}
    \Omega = \Omega_0 + \nu \,, \\
    J = \frac{\Omega_0^3(1-\eta^2)}{\sqrt 2}
  \end{gather}
\end{subequations}
where $|\nu|\ll 1$, $\eta^2 \ll 1$.

In this approximation,
\begin{subequations}
  \begin{align}
    P_1 &\approx \Omega_0^4\left(
      4\nu^2-\eta^2\Omega_0^2
      - \frac{\gamma_0\Omega_0}{\sqrt{2}}\sin2\alpha
    \right)^2 \,, \\
    P_2 &\approx 2\gamma_0^2\Omega_0^6(1+\cos^2\alpha)^2\,, \\
    P_3 &\approx |{\cal A}_0|^2\Omega_0^4
      \left[(4\nu^2-\nu^2\Omega_0^2)^2 + 4\gamma_0^2\Omega_0^2\right]\,,\\
    Q &\approx 2\sqrt{2}\gamma_0\Omega_0^7(1+\cos^2\alpha)\,,
  \end{align}
\end{subequations}
and
\begin{subequations}
  \begin{align}
    Z &\approx m(4\nu^2-\eta^2\Omega_0^2-2i\Omega_0\gamma_0)\,,
      \label{Z_approx}\\
    Z_{\rm eff}& \approx m\left(
       4\nu^2-\eta^2\Omega_0^2
       - \frac{\Omega_0\gamma_0}{\sqrt{2}}\sin2\alpha
     \right) \,, \label{Z_approx_eff} \\
   S_x &\approx \frac{\hbar}{\sqrt{2}m\gamma_0\Omega_0(1+\cos^2\alpha)}\,.
  \end{align}
\end{subequations}
In this approximation we have the following formula for the sensitivity
$\xi$:
\begin{multline}\label{xi2nb}
  \xi^2(\nu) \approx
    \frac{\gamma_0}{\Omega_0}\,\frac{(1+\cos^2\alpha)}{\sqrt{2}}
    + \frac{1}{2\sqrt{2}\gamma_0\Omega_0^3(1+\cos^2\alpha)} \\
      \times\biggl\{
        \left(
          4\nu^2-\eta^2\Omega_0^2
          - \dfrac{\gamma_0\Omega_0}{\sqrt{2}}\sin 2\alpha
         \right)^2 \\
         + |{\cal A}_0|^2
           \left[
             \left(4\nu^2-\eta^2\Omega_0^2\right)^2
             + 4\Omega_0^2\gamma_0^2
           \right]
      \biggr\}\,.
\end{multline}
Using the following notations:
\begin{subequations}\label{eta_A_alpha}
  \begin{gather}
    \eta_\alpha^2 = \eta^2 + \dfrac{\gamma_0\sin 2\alpha}
      {\sqrt{2}\Omega_0(1+|{\cal A}_0|^2)}\,,\\
    |{\cal A}_\alpha|^2 = \frac{\sqrt{2}|{\cal A}_0|^2}{1+\cos^2\alpha} \,,
  \end{gather}
\end{subequations}
and taking into account that
\begin{equation}
  \gamma_0=\gamma_0^\text{loss}\left(1+\frac{1}{|{\cal A}_0|^2}\right)\,,
\end{equation}
Eq.\,(\ref{xi2nb}) can be rewritten in a more compact form:
\begin{equation}
\label{xi(nu)}
  \xi^2(\nu) \approx
    \frac{\left(4\nu^2-\eta_\alpha^2\Omega_0^2\right)^2}{4g\Omega_0^4}
    + gC \,,
\end{equation}
where
\begin{gather}
  g = \frac{\gamma_0^{\rm loss}}{\Omega_0|{\cal A}_\alpha|^2} \,, \\
  C = 1 + \frac{3}{\sqrt{2}}\,|{\cal A}_\alpha|^2 + |{\cal A}_\alpha|^4 \,.
\end{gather}

\subsection{Signal-to-noise ratio}\label{app:analysis:snr}

\begin{figure}[t]
  \psfrag{x}{$|{\cal A}_\alpha|$}
  \psfrag{snr}[rc][rc]{$k(|{\cal A}_\alpha|)$}
  \includegraphics[width=0.44\textwidth,height=0.33\textwidth]{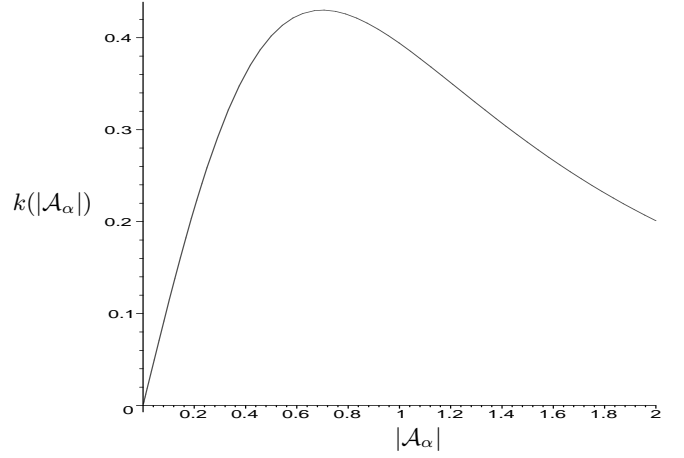}
  \caption{Dependence of the signal-to-noise ratio on
  $|{\cal A}_\alpha|$ at optimal $\eta_\alpha^2$ (\ref{eta_opt}).}
  \label{fig:snrdbl}
\end{figure}

In the narrow-band case, the main contribution into the integral of
the signal-to-noise ratio is produced in vicinity of $\Omega_0$. In this
case we can use approximation (\ref{xi(nu)}) and expand the limits of
integration over $\nu$ from $-\infty$ to $\infty$ and making substitution
$x=\sqrt 2\nu/\Omega_0\sqrt g\, C^{1/4}$ with notation
$a=\eta^2_{\alpha}/2g\sqrt C$:
\begin{multline}
  {\rm SNR} = \frac{2}{\pi}\int_0^{\infty}
    \frac{|h(\Omega)|^2\,d\Omega}{S_h(\Omega)}
  \approx
    \frac{2}{\pi}\,\frac{|h(\Omega_0)|^2}{h_{\rm SQL}^2(\Omega_0)}
    \int_{-\infty}^\infty\frac{d\nu}{\xi^2(\nu)} \\
  = \frac{2|h(\Omega)|^2\Omega_0}{\pi h_{\rm SQL}^2(\Omega_0)} \times
    \frac{1}{\sqrt{2g}C^{3/4}} \int_{-\infty}^\infty
    \frac{dx}{1+(x^2+a^2)^2}\\
  =\frac{|h(\Omega)|^2\Omega_0}{ h_{\rm SQL}^2(\Omega_0)} \times
    \frac{1}{\sqrt{2g}\, C^{3/4}}\left(
    \frac{1}{\sqrt{a+i}} +
    \frac{1}{\sqrt{a-i}}\right)=\\
  = \frac{|h(\Omega)|^2\Omega_0}{ h_{\rm SQL}^2(\Omega_0)} \times
    \frac{1}{\sqrt{2g}\, C^{3/4}}\times
    \frac{\sqrt{a+\sqrt{a^2+1}}}{\sqrt 2\sqrt{a^2+1}}
\end{multline}
The maximum of this expression is achieved, if
\begin{equation}
\label{eta_opt}
 a=\frac{1}{\sqrt 3},\quad \text{or}\quad
 \eta_\alpha^2 = \frac{2g\sqrt{C}}{\sqrt{3}}\,,
\end{equation}
and it is equal to:
\begin{equation}
  {\rm SNR} = k(|{\cal A}_\alpha|)\times
    \frac{|h(\Omega)|^2\Omega_0}{h_{\rm SQL}^2(\Omega_0)}\,
    \sqrt{\frac{\Omega_0}{\gamma_0^{\rm loss}}} \,,
\end{equation}
where
\begin{equation}
  k(|{\cal A}_\alpha|) = \frac{3^{3/4}|{\cal A}_\alpha|}{2C^{3/4}}
\end{equation}
is dimensionless function plotted in Fig.\,\ref{fig:snrdbl}. It attains the maximum
at
\begin{equation}
  |{\cal A}_\alpha| = \sqrt{\frac{\sqrt{73}-3}{8\sqrt{2}}} \approx 0.61
\end{equation}
and it is equal to
\begin{equation}
  \frac{2^{7/4}\sqrt{\sqrt{73}-3}}{(23+3\sqrt{73})^{3/4}}
  \approx 0.43 \,.
\end{equation}

\section{Gain in spectral density for simplified model without optical
losses}\label{app:xi}

Starting with this Appendix we simplify the model system. In particular, we assume here, that there is no optical losses
in the system we examine. When considering the frequency-dependent rigidity
based system, the approximate formula (\ref{K_simple}) is used.

\subsection{Conventional harmonic oscillator}

Spectral density of the total meter noise (\ref{F_noise}) is equal to:
\begin{equation}\label{app_S_sum}
  S_{\rm sum}(\Omega) = S_F(\Omega) + |Z(\Omega)|^2S_x(\Omega) \,,
\end{equation}
where $Z(\Omega)$ is the spectral image of the differential operator ${\bf
Z}$. In case of a conventional harmonic oscillator,
\begin{equation}\label{app_Z_oscill}
  |Z(\Omega)|^2 = (-m\Omega^2 + K)^2 \,,
\end{equation}
and in close vicinity of the resonance frequency
$\Omega_0=\sqrt{K/m}$,
\begin{equation}\label{app_Z_oscill_nu}
  |Z(\Omega)|^2 \approx 4m^2\Omega_0^2\nu^2 \,,
\end{equation}
and
\begin{equation}
  \xi^2(\Omega)\equiv\frac{S_{\rm sum}(\Omega)}{2\hbar m\Omega^2}
    \approx \xi_{\rm min}^2 + \dfrac{\nu^2}{\xi_{\rm min}^2\Omega_0^2} \,,
\end{equation}
where
\begin{gather}
  \nu = \Omega-\Omega_0 \,, \quad |\nu| \ll \Omega_0 \,,\label{def_nu}\\
  \xi_{\rm min}^2 = \frac{S_F(\Omega_0)}{2\hbar m\Omega_0^2} \,,
\end{gather}
see Eqs\,(\ref{SxSF}, \ref{S_SQL}).

Let us require that $\xi(\Omega)$ does not exceed a given value $\xi_0$ within as
wide a frequency band $\Delta\Omega$ as possible. It is easy to show that this
requirement is met if
\begin{equation}
  \xi_{\rm min} = \frac{\xi_0}{\sqrt{2}} \,,
\end{equation}
and
\begin{equation}
  \Delta\Omega = 2\xi_{\rm min}^2\Omega_0 \,.
\end{equation}
Therefore,
\begin{equation}
  \xi_0^2 = \frac{\Delta\Omega}{\Omega_0} \,.
\end{equation}

\subsection{Frequency-dependent rigidity}

\subsubsection{Double-resonance case}

Consider now an oscillator with the frequency-dependent rigidity
(\ref{K_simple}). In this case,
\begin{equation}
  |Z(\Omega)|^2 = \frac
    {m^2(\Omega_+^2 - \Omega^2)^2(\Omega_-^2 - \Omega^2)^2}
    {(\Omega_+^2 + \Omega_-^2 -\Omega^2)^2}
\end{equation}
[see Eq.\,(\ref{Omega_pm})].

If the double-resonance condition (\ref{dbl_res}) is fulfilled, then in
close vicinity of the resonance frequency
$\Omega_0=\delta_0/\sqrt{2}$,
\begin{gather}
  |Z(\Omega)|^2 \approx 16m^2\nu^4 \,, \\
  \xi^2(\Omega) \approx \xi_{\rm min}^2
    + \dfrac{4\nu^4}{\xi_{\rm min}^2\Omega_0^4} \,.
\end{gather}
Performing again the same optimization as in previous subsection, we can
obtain that again $\xi_{\rm min} = \xi_0/\sqrt{2}$, and
\begin{equation}
  \xi_0 = \frac{\Delta\Omega}{\Omega_0} \,.
\end{equation}

\subsubsection{Two close resonances}

In the sub critical pumping case (\ref{E_subcrit}),
\begin{gather}
  |Z(\Omega)|^2 \approx m^2(4\nu^2-\Omega_0^2\eta^2)^2\,,\label{Z_fdrigid}\\
  \xi^2(\Omega) \approx \xi_{\rm min}^2
    + \dfrac{(4\nu^2-\Omega_0^2\eta^2)^2}{4\xi_{\rm min}^2\Omega_0^4} \,.
\end{gather}
These functions has a local maximum at $\nu=0$ and two minima at
$\nu=\pm\Omega_0\eta/2$.

The same optimization as in two previous cases gives, that
\begin{gather}
  \xi(\Omega_0) = \xi_0 = \sqrt{2}\xi_{\rm min} \,, \\
  \xi_0 = \frac{1}{\sqrt{2}}\,\frac{\Delta\Omega}{\Omega_0} \,,
\end{gather}
and the optimal value of parameter $\eta$ is equal to
\begin{equation}
  \eta_c = \xi_0\,,
\end{equation}

\section{Gain in signal-to-noise ratio for simplified model
    without optical losses}\label{app:snr}

\subsection{Free test masses interferometer}

Rewrite the signal-to-noise ratio (\ref{snr}) as follows:
\begin{equation}\label{snrF}
  {\rm SNR} = \frac{2}{\pi}\int_0^\infty
    \frac{|F_{\rm signal}(\Omega)|^2\,d\Omega}{S_{\rm sum}(\Omega)} \,.
\end{equation}
For conventional interferometer (without optical springs), the total meter noise
spectral density is equal to (see \cite{02a1KiLeMaThVy}):
\begin{equation}
  S_{\rm sum}(\Omega) = \frac{\hbar m}{2}\left[
    \frac{2\Omega_0^4}{\Omega_0^2+\Omega^2}
    + \frac{\Omega^4(\Omega_0^2+\Omega^2)}{2\Omega_0^4}
  \right] \,.
\end{equation}
Therefore, in this case,
\begin{multline}
  {\rm SNR} = \frac{8}{\pi\Omega_0^2h_{\rm SQL}^2(\Omega_0)}
    \int_0^\infty
      \frac{|h(\Omega)|^2\,d\Omega}{
        \dfrac{2\Omega_0^4}{\Omega^4(\Omega_0^2+\Omega^2)}
        + \dfrac{\Omega_0^2+\Omega^2}{2\Omega_0^4}
      } \\
  = {\cal N}\times\frac{|h(\Omega_0)|^2\Omega_0}{h^2_{SQL}(\Omega_0)} \,,
\end{multline}
where
\begin{equation}
  {\cal N} = \frac{8}{\pi\Omega_0^3|h(\Omega_0)|^2}
    \int_0^\infty
      \frac{|h(\Omega)|^2\,d\Omega}{
        \dfrac{2\Omega_0^4}{\Omega^4(\Omega_0^2+\Omega^2)}
        + \dfrac{\Omega_0^2+\Omega^2}{2\Omega_0^4}
      } \,.
\end{equation}
is the numeric factor depending on the gravitation-wave signal shape
$h(\Omega)/|h(\Omega_0)|$.

\subsection{Conventional harmonic oscillator}

Substituting Eqs.\,(\ref{app_S_sum}) and (\ref{app_Z_oscill}) into
Eq.\,(\ref{snrF}), we obtain, that for a conventional harmonic
oscillator the  signal-to-noise ratio is equal to:
\begin{equation}
  {\rm SNR} = \frac{2}{\pi}
    \int_0^\infty\frac{|F_{\rm signal}(\Omega_0)|^2\,d\Omega}
      {S_F(\Omega) + m^2S_x(\Omega)(\Omega_0^2-\Omega^2)^2} \,.
\end{equation}
In the narrow-band case [see Eqs.\,(\ref{SxSF0}, \ref{def_nu})], this
equation can be presented as the following:
\begin{multline}
  {\rm SNR} \approx \frac{2}{\pi}|F_{\rm signal}(\Omega_0)|^2
    \int_{-\infty}^\infty\frac{d\nu}
      {S_F(\Omega_0) + 4m^2\Omega_0^2S_x(\Omega_0)\nu^2} \\
  = \frac{|F_{\rm signal}(\Omega_0)|^2}{\hbar m\Omega_0}
  = \frac{2|h(\Omega_0)|^2\Omega_0}{h_{\rm SQL}^2(\Omega_0)} \,.
\end{multline}

\subsection{Frequency-dependent rigidity}

In similar way, using Eqs.\,(\ref{app_S_sum},\ref{Z_fdrigid}) and
(\ref{snrF}), we obtain, that for optical spring based
oscillator,
\begin{multline}
  {\rm SNR}
  \approx \frac{2}{\pi}|F_{\rm signal}(\Omega_0)|^2 \\
    \times\int_{-\infty}^\infty\frac{d\nu}
      {S_F(\Omega_0) + m^2S_x(\Omega_0)(4\nu^2-\Omega_0^2\eta^2)^2} \\
  = \frac{\sqrt{2}|h(\Omega_0)|^2\Omega_0}{h_{\rm SQL}^2(\Omega_0)}\,
      \frac{\xi_0^2}{\sqrt{
        \bigl(\xi_0^4+\eta^4\bigr)\bigl(\sqrt{\xi_0^4+\eta^4}-\eta^2\bigr)
      }}\,.
\end{multline}

In a pure double resonance case ($\eta=0$),
\begin{equation}
  {\rm SNR} = \frac{\sqrt{2}}{\xi_0}\,
    \frac{|h(\Omega_0)|^2\Omega_0}{h_{\rm SQL}^2(\Omega_0)}\,.
\end{equation}
Slightly better result can be obtained for the case of two optimally
placed resonances. If
\begin{equation}
  \eta^2 = \frac{\eta_c^2}{\sqrt{3}} \equiv \frac{\xi_0^2}{\sqrt{3}}\,,
\end{equation}
then
\begin{equation}
  {\rm SNR} = \frac{3^{3/4}}{\sqrt{2}\xi_0}
    \frac{|h(\Omega_0)|^2\Omega_0}{h_{\rm SQL}^2(\Omega_0)}\,,
\end{equation}

If the separation between the two resonance frequencies is too big,
$\eta\gg \xi_0$ (but still $\eta\ll 1$), then
\begin{equation}
  {\rm SNR} = \frac{2}{\eta}\,
    \frac{|h(\Omega_0)|^2\Omega_0}{h_{\rm SQL}^2(\Omega_0)}\,.
\end{equation}



\end{document}